\documentclass[a4paper,USenglish]{article}

\usepackage{amsmath}
\usepackage{amsthm}
\usepackage{amssymb}
\usepackage{enumerate}
\usepackage{tikz}
\usepackage[colorlinks]{hyperref}

\definecolor{lightblue}{rgb}{0.5,0.5,1.0}
\definecolor{darkred}{rgb}{0.5,0,0}
\definecolor{darkgreen}{rgb}{0,0.5,0}
\definecolor{darkblue}{rgb}{0,0,0.5}

\hypersetup{colorlinks,linkcolor=darkred,filecolor=darkgreen,urlcolor=darkred,citecolor=darkblue}

\newtheorem{theorem}{Theorem}
\newtheorem{lemma}[theorem]{Lemma}
\newtheorem{corollary}[theorem]{Corollary}

\newtheorem{definition}[theorem]{Definition}
\theoremstyle{definition}

\theoremstyle{remark}

\newtheorem*{note*}{Note}

\newtheorem*{remark*}{Remark}

\newcommand{\supp}{\mathrm{supp}}
\newcommand{\Fp}{\mathbb{F}_p}
\newcommand{\lv}{\vartheta}
\newcommand{\diag}[1]{\mathrm{diag}(#1)}
\newcommand{\binary}{\mathbb{B}}
\newcommand{\isom}{\cong}
\newcommand{\symmdif}{\triangle}

\usepackage{mathtools}

\title{Resolution with Symmetry Rule applied to Linear Equations\footnote{The research leading to these results has received funding from the European Research Council (ERC) under the European Union’s Horizon 2020 research and innovation programme (EngageS: grant agreement No. 820148).}}
\author{Pascal Schweitzer \and Constantin Seebach\\}
\date{TU Kaiserslautern} 

\usepackage{fullpage}

\begin{document}
\maketitle
\thispagestyle{empty}
\begin{abstract}

This paper considers the length of resolution proofs when using Krishnamurthy's classic symmetry rules. We show that inconsistent linear equation systems of bounded width over a fixed finite field~$\Fp$ with~$p$ a prime have, in their standard encoding as CNFs, polynomial length resolutions when using the local symmetry rule (SRC-II).

As a consequence it follows that the multipede instances for the graph isomorphism problem encoded as CNF formula have polynomial length resolution proofs. This contrasts exponential lower bounds for individualization-refinement algorithms on these graphs.

For the Cai-Fürer-Immerman graphs, for which Tor{\'{a}}n showed exponential lower bounds for resolution proofs (SAT 2013), we also show that already the global symmetry rule (SRC-I) suffices to allow for polynomial length proofs.
\end{abstract}

\clearpage
\setcounter{page}{1}

\section{Introduction}

Refutation via logical resolution is one of the most basic and fundamental methods in theorem proving used to argue the validity of statements in propositional logic.
It is famously sound and complete for proving that formulas in conjunctive normal form (CNF) are unsatisfiable. In automated theorem proving, resolution is in particular used for various primitive backtracking algorithms for the satisfiability problem (SAT) such as the DPLL algorithm.

However, resolution is primitive in that we know simple unsatisfiable CNF formulas that admit only resolution refutations of superpolynomial length. This was first proven by Haken~\cite{DBLP:journals/tcs/Haken85} who showed that a canonical encoding of the pigeonhole principle into a CNF formula provides formulas whose shortest refutations are superpolynomial in length. Other examples and exponential bounds were given by Chv{\'{a}}tal and Szemer{\'{e}}di~\cite{DBLP:journals/jacm/ChvatalS88} as well as Urquhart who used formulas based on Tseitin tautologies~\cite{UrquhartHardRes}.
Investigating the resolution complexity of the graph non-isomorphism problem, Tor{\'{a}}n~\cite{Toran2013} constructed 
CNF formulas from so-called CFI-graphs (see~\cite{DBLP:journals/combinatorica/CaiFI92}) and showed  the shortest resolution proofs of the arising formulas have exponential length.

As observed by Krishnamurthy, many simple examples without short resolution refutations exhibit symmetries. This prompted the introduction of Krishnamurthy's symmetry rule~\cite{Krishnamurthy1985} which intuitively allows the deduction of a clause symmetric to a previously deduced clause in one step (formal definitions are given in Section~\autoref{sec:prelims}). For various formulas, Krishnamurthy argued polynomial bounds when the symmetry-rule is used, leading to exponential improvements. Further examples with this effect, including another analysis for pigeonhole principle formulas, were provided by Urquhart~\cite{URQUHART1999}. 

Krishnamurthy in fact introduced two rules, each of them arises from permutations of the variables. The \emph{global} rule allows only symmetries of the entire original formula, while the \emph{local} one allows us to use symmetries of a subset of the clauses. These rules led to the proof systems SR-I (\emph{symmetric resolution}) and SR-II (\emph{locally symmetric resolution}), respectively. Urquhart~\cite{URQUHART1999} introduced \emph{complementation symmetries} in addition to the variable permutations. This allows us to interchange literals with their negations and leads to the proof systems SRC-I and SRC-II. In~\cite{URQUHART1999} Urquhart also showed that there are exponential-to-polynomial improvements regarding proof length from the system SR-I to SRC-I. Arai and Urquhart~\cite{AraiUrquhart} showed exponential-to-polynomial improvements from SR-I to SR-II and also provided exponential lower bounds for SRC-II.

Szeider~\cite{Szeider2005}, who actually focuses on homomorphisms, describes another strengthening of the symmetry rule. In his extension we are allowed the use of symmetries within clauses that have been resolved, rather than only allowing clauses of the original formula. This is called resolution with \emph{dynamic} symmetries and leads to the proof systems SR-III and SRC-III, depending on whether complementation is allowed. However, to date it remains an open problem to find superpolynomial lower bounds on proof length in SR-III and SRC-III.

\subsection{Contribution.} In this paper we are concerned with proof systems obtained by extending resolution with additional symmetry rules. We prove that the CNF formulas arising from the CFI-graphs have refutations polynomially bounded in length in the SR-I calculus. With Tor{\'{a}}n's exponential lower bounds~\cite{Toran2013} mentioned above, this gives an exponential-to-polynomial improvement for the resolution complexity of non-isomorphism when introducing the symmetry rule.
To those familiar with the details of the CFI-construction this may not come as a surprise, since the CFI-graphs exhibit many global symmetries. However, this is not the case for multipede graphs, these arise from a construction related to the CFI-graphs~\cite{DBLP:journals/jsyml/GurevichS96}. Crucially these graphs are asymmetric. That is, they have no symmetries at all. They provide exponential lower bounds for all individualization-refinement algorithms for the graph isomorphism problem. This includes all tools currently viable in practice, such as nauty/traces~\cite{McKayPiperno}. The initial intuition might therefore be that the CNF formulas arising from multipedes provide exponential lower bounds for SRC-III. However, this turns out not to be the case. In fact, maybe surprisingly, we show that even when using only local symmetries rather than dynamic symmetries (i.e., in SCR-II rather than SCR-III) there are polynomial bounds on the respective formulas. In some sense this shows that the multipedes have substructures with symmetries that allow them to be distinguished concisely.

To prove this statement, we reduce the statement to one concerning linear equation systems. It is known that isomorphism of CFI and multipede graphs are related to solvability of linear equation systems. (This is also the case for Tseitin tautologies.) We show that this relation can be exploited.  Specifically, we show that there is a resolution transforming the CNFs arising from the graph isomorphism instances to CNFs arising from linear equation systems. We then show our main theorem which says that inconsistent linear equation systems with equations of bounded width (i.e.,~the maximum number of non-zero coefficients in an equation is bounded) have polynomial resolutions using the local symmetry rule.

\begin{theorem}
Inconsistent linear equation systems of bounded width over a fixed finite field~$\Fp$ with~$p$ a prime have, in their standard encoding as CNFs, polynomial length resolutions when using the local symmetry rule (i.e., in SRC-II).
\end{theorem}

\textbf{Structure of the paper.} 

\autoref{sec:cfi res} shows that the CNF formulas arising from CFI-graph pairs have polynomial length proofs in SR-I. 
\autoref{sec:lin:eqs} shows that linear equation systems of bounded width have polynomial length proofs in SRC-II.
\autoref{sec:multipedes} shows that the formulas arising from (bounded degree) multipede graphs can be transformed in the resolution calculus (without using symmetry) to linear equation systems of bounded width.

\subsection{Related Work} \autoref{systems:overview} gives an overview of resolution calculi with symmetry and references to lower bound constructions.
A proof system \emph{p-simulates} another proof system if shortest proofs in the latter are polynomially bounded in the length of shortest proofs in the former. We should remark that the extended resolution system introduced by Tseitin~\cite{Tseitin} can p-simulate proof systems with symmetries~\cite{URQUHART1999}.
See~\cite{DBLP:journals/jar/BenhamouS94} for an implementation using Krishnamurthy's symmetry rule. 
Symmetry rules have of course also been introduced for other proof systems~\cite{DBLP:conf/sat/BlinkhornB19,DBLP:conf/lpar/Egly93}. See also~\cite{DBLP:conf/aaai/DixonGHLP04} for another way to incorporate symmetries into resolution.

\begin{figure}
\centering
\begin{tabular}{r|c|c|c|c}

 & none  &  global & local  & dynamic \\ 
\hline 
without complementation & classical resol.\cite{DBLP:journals/jacm/ChvatalS88,UrquhartHardRes} & SR-I \cite{URQUHART1999} & SR-II~\cite{AraiUrquhart} &  SR-III (open)  \\ 
\hline 
with complementation   & -- & SRC-I \cite{URQUHART1999} & SRC-II~\cite{AraiUrquhart} &  SRC-III (open)  \\ 
\end{tabular} 
\caption{Resolution calculi with symmetries rules of varying degree of generality and references with formulas proving exponential lower bounds on resolution length.}\label{systems:overview}
\end{figure}

\textbf{Connection to the graph isomorphism problem.} The results of our paper are connected to the graph isomorphism problem in two conceptually very different ways. First, finding valid literal permutations (with or without complementation) for the global symmetry rule is equivalent to the graph isomorphism problem itself (e.g.,~\cite{DBLP:journals/combinatorica/CaiFI92}). Therefore isomorphism solvers such as nauty/traces~\cite{McKayPiperno}, which are highly efficient in practice, can be used to find the symmetries (see~\cite{DBLP:conf/ijcai/TourD95}). 
Symmetry detection is one of the standard applications of graph isomorphism solvers, for example there is a tool integrating nauty into Prolog~\cite{FrankC16} for this purpose.

Second, our results relate to the proof complexity of the graph isomorphism problem itself, which explains why we are interested in CNF formulas arising from non-isomorphism instances. Tor{\'{a}}n~\cite{Toran2013} describes a canonical way to encode the isomorphism problem as a CNF formula (see \autoref{subsec:encoding:griso}). 
The resolution complexity of graph non-isomorphism is related to the complexity of the graph isomorphism problem.
After all isomorphism solvers need to prove, some way or another, that the inputs are non-isomorphic, if they are. A crucial feature of isomorphism solvers is that they are able to exploit already detected symmetries (i.e., automorphisms) of the underlying instances during run-time~\cite{McKayPiperno}. Vaguely, this translates into a symmetry rule that they apply already during the process of computing the symmetries of the instance. Current tools basically only exploit local symmetries. Our new insights into the resolution complexity of multipedes thus shows a combinatorial possibility  to solve their isomorphism problem. It brings up the question how to exploit local symmetries in graph isomorphism solvers.

It remains unknown whether graph non-isomorphism has polynomial resolution complexity in any of the proof systems with symmetry rule we have discussed.

\section{Preliminaries}\label{sec:prelims}
\subsection{Resolution and the Symmetry Rule}
We are interested in unsatisfiability proofs of Boolean formulas. The basic resolution proof system works with formulas in conjunctive normal form.

	Let $\Gamma$ be a finite set of variables.
		$\mathrm{Lit}(\Gamma) := \Gamma \cup \overline{\Gamma}$ is the set of literals, where $\overline{\Gamma} := \{ \overline{x} \mid x \in \Gamma \}$.
		A \emph{clause} is a disjunction of literals. We also represent clauses as sets of literals. 
		A Boolean formula is in \emph{conjunctive normal form (CNF)} if it is a conjunction of clauses. We may treat such a formula as a set of clauses.
		$\bot$ is the empty clause, i.e. the disjunction of the empty set, which is unsatisfiable.
		For sets of clauses $C_1$ and $C_2$ define $C_1 \sqsubseteq C_2 :\iff \forall c_1 \in C_1 \exists c_2\in C_2: c_1\supseteq c_2$.
Since we will treat clauses as sets of literals, we do not care for their order, i.e.\ we do not differentiate between $x \vee y$ and $y \vee x$. The same applies to CNF formulas, which we interpret as sets of clauses.

\begin{definition}
	\emph{Resolution} is a proof system in propositional logic. It operates on CNF formulas, employing a single inference rule: \[\frac{x\vee A, \; \overline{x} \vee B}{A \vee B}.\] The clause produced by the resolution rule is called \emph{resolvent}.
	
	Let $A=\{a_1,\dots,a_m\}$ and $B$ be sets of clauses. We write $A \vdash_n B$ if there exists a sequence of clauses $a_1,\dots,a_m,c_1,\dots,c_n$ such that every $c_i$ is a resolvent of two earlier clauses and $B \subseteq A \cup\{c_1,\dots,c_n\}$. Such a sequence is called \emph{derivation} of $B$ from $A$.
	When the length of the sequence is irrelevant, we write $A \vdash B$, meaning $A \vdash_n B$ for some $n$.
	Given a clause $b$, we also write $A \vdash_n b$ for $A \vdash_n \{b\}$.
	
	For a CNF formula $F$ with $F\vdash_n \bot$, we say $F$ has a \emph{resolution refutation} of size $n$.
	
	We write $A\vdash^w_n B$ if there exists a set of clauses $B'$ such that $B \sqsubseteq B'$ and $A\vdash_n B'$. This is a weaker requirement than $A\vdash_n B$.
\end{definition}
Resolution is sound and complete, i.e. $F\vdash_n \bot$ if and only if $F$ is unsatisfiable.
We examine the proof complexity of formulas in this proof system, i.e., the length of the shortest possible resolution refutation of a given formula, in relation to the formula size. 
There exist classes of formulas with exponential lower bounds on the resolution proof complexity \cite{DBLP:journals/jacm/ChvatalS88, Toran2013, UrquhartHardRes}.

In the following we define the symmetry rule, which is an extension to resolution, aiming to reduce the proof complexity of some of these hard formulas.

\begin{definition}
	Let $L$ be a finite set of literals. A bijection $\sigma: L \rightarrow L$ is called \emph{renaming} if for every $\ell \in L$ we have $\overline{\sigma(\ell)} = \sigma(\overline{\ell})$.
\end{definition}
A renaming is essentially a permutation of the variables that may also negate some of them. We can apply renamings to clauses (i.e., sets of literals) and CNF formulas (i.e., sets of clauses). In either case we define $\sigma(C) := \{\sigma(x) \mid x \in C\}$.

\begin{definition}[The Symmetry Rule \cite{AraiUrquhart} \cite{Szeider2005}] 
	Consider a derivation $S$ from a formula $F$ and a subsequence $S'$ of $S$ which derives a clause $C$ from a subset $F' \subseteq F$.
	If there exists a renaming $\sigma$ with $\sigma(F') \subseteq F$, then the \emph{local symmetry rule} allows derivation of $\sigma(C)$.
	
	With the restriction $F=F'$, we obtain the \emph{global symmetry rule}.
	Adding the global or local symmetry rule to the resolution system yields the proof systems \textbf{SRC-I} and \textbf{SRC-II}, respectively.
\end{definition}
\newcommand{\lgsymd}{\vdash^{\text{SRC-I}}}
\newcommand{\lsymd}{\vdash^{\text{SRC-II}}}

We write $A \lsymd_n B$ to indicate that $B$ can be derived from $A$ using resolution and the local symmetry rule, with a derivation of length at most $n$.

Note that in order to apply $\sigma$ via the local symmetry rule to some clause $C$ in a derivation, we must look at the entire history of how $C$ was derived, and find out which part $F' \subseteq F$ of the original formula was used. Then we need to check that $\sigma(F') \subseteq F$.

This means that in general we cannot chain derivations that use the symmetry rule together, because such an operation changes the history for some of the clauses. Still, we can combine \textbf{SRC-II} derivations in the following ways:

\begin{lemma} \label{reshelper1}
Let $A,B,C$ and $D$ be sets of clauses and $n,m\in \mathbb{N}$.
\begin{alphaenumerate}
	\item $A \lsymd_n B$ and $A \subseteq C$ implies $C \lsymd_n B$
	\item $A \lsymd_n B$ and $B \vdash_m C$ implies $A \lsymd_{n+m} C$
	\item $A \lsymd_n B$ and $C \lsymd_m D$ implies $A\cup C \lsymd_{n+m} B\cup D$
\end{alphaenumerate}
\end{lemma}

\begin{lemma} \label{reshelper2}
	Let $A$ and $B$ be sets of clauses and $d$ a clause.
	\begin{alphaenumerate}
		\item $A \vdash \bot$ and $\{c \vee d \mid c\in A\} \sqsubseteq B$ implies $B \vdash^w d$
	\end{alphaenumerate}
\end{lemma}

\newcommand{\edgeset}[1]{\mathcal{E}_{#1}}
\newcommand{\neighborhood}[1]{N_{#1}}
\subsection{Encoding Graph Isomorphism}\label{subsec:encoding:griso}
Our interest in the graph isomorphism problem is twofold: First, finding valid literal permutations for the symmetry rule is equivalent to finding certain graph isomorphisms. Secondly, we examine the proof complexity of the problem by translating it into propositional logic and applying resolution with symmetry rule.

		A \emph{graph} is a tuple $(V,E)$ of a set of vertices $V$ and edges $E$. Each edge is a two element subset of $V$.
		A \emph{colored graph} is a graph $(V,E)$ together with a function $f:V\rightarrow C$, called \emph{coloring}, assigning to every vertex a color from some set $C$.
	Let $G=(V,E)$ be a graph and $v \in V$.
		$\edgeset{G}(v) := \{ e \in E \mid v\in e \}$ are the edges \emph{incident} with $v$.
		$\neighborhood{G}(v) := \{ u \in V \mid \{u,v\}\in E \}$ is the \emph{neighborhood} of $v$.
		$\deg_{G}(v) := |\neighborhood{G}(v)| = |\edgeset{G}(v)|$ is the \emph{degree} of $v$.

Given a colored graph $G=(V,E)$ with coloring $f$ and a vertex $v\in V$, we can \emph{individualize} $v$ by creating a new coloring $f'$ such that $f'(v):= (f(v), 1)$ and setting  $f'(v'):= (f(v'), 0)$ for all $v'\in V\setminus \{v\}$. We write the individualized graph as $G_v$.

\begin{definition}
	Let $G_1=(V_1,E_1)$ and $G_2=(V_2,E_2)$ be graphs.
	\begin{itemize}
		\item A \emph{graph isomorphism} from $G_1$ to $G_2$ is a bijection $\varphi: V_1\rightarrow V_2$ such that for all $v,v'\in V_1$ we have $\{v,v'\}\in E_1$ if and only if $\{\varphi(v), \varphi(v')\}\in E_2$.
		\item We say $G_1$ and $G_2$ are \emph{isomorphic}, written $G_1 \isom G_2$, if there exists a graph isomorphism from $G_1$ to $G_2$.
		\item An \emph{automorphism} of a graph $G$ is a graph isomorphism from $G$ to itself.
		\item $\mathrm{Aut}(G)$ is the \emph{automorphism group} of G.
	\end{itemize}
\end{definition}
The automorphisms of a graph constitute its inherent combinatorial symmetries.
We will use the terms automorphism and symmetry synonymously.

Given two graphs $G_1$ and $G_2$, one can construct a Boolean formula that is satisfiable if and only if there is an isomorphism between $G_1$ and $G_2$ \cite{Toran2013}. This is commonly done by constraining variables of the form $x_{u,v}$ such that each satisfying assignment corresponds to an isomorphism: $1$ is assigned to $x_{u,v}$ if and only if the isomorphism maps $u$ to $v$.

\newcommand{\xor}{\oplus}
\begin{definition} \label{isomorphism formula}
	For a pair of graphs $G_1=(V_1,E_1), G_2=(V_2,E_2)$ with $|V_1|=|V_2|$, define
	\begin{align*}
	F(G_1,G_2) &:= T_1 \wedge T_2 \wedge T_3, \text{ where} \\
	T_1 &:= \bigwedge_{v_1 \in V_1} \bigvee_{v_2 \in V_2} x_{v_1,v_2} ,\\
	T_2 &:= \bigwedge_{v_2 \in V_2}\bigwedge_{\substack{v_1,v_1' \in V_1 \\ v_1\neq v_1'}} \left( \overline{x_{v_1,v_2}} \vee \overline{x_{v_1',v_2}} \right) ,\\
	T_3 &:= \bigwedge_{\{u_1,v_1\} \in E_1 \xor \{u_2,v_2\} \in E_2} \left( \overline{x_{u_1,u_2}} \vee \overline{x_{v_1,v_2}} \right) .
	\end{align*}\end{definition}
	We refer to the clauses of this CNF formula as being \emph{``of Type i''}, depending on which $T_i$ they come from.
The clause types naturally encode the concept of a graph isomorphism in propositional logic. Specifically, Type 1 and Type 2 clauses ensure that we have a bijection from $V_1$ to $V_2$; Type 3 clauses make the function preserve edges.

If the graphs $G_1$ and $G_2$ are colored by some functions $l_1$ and $l_2$ respectively, then an isomorphism between them should respect the colors. To represent this in the formula $F(G_1,G_2)$, we simply assign $0$ to all variables $x_{u,v}$ for which $l_1(u)\neq l_2(v)$.

\subsection{The CFI Graphs}
\begin{figure}
	\centering\scalebox{0.8}{
	\begin{tikzpicture}[scale=0.8,semithick]
	\tikzstyle{every node}=[draw,circle]
	
	\node[fill=gray] (m) at (-3,0) [label=right:{$m_\emptyset$}]{};
	\node[fill=gray] (m12) at (-1,0) [label=right:{$m_{\{1,2\}}$}]{};
	\node[fill=gray] (m13) at (1,0) [label=right:{$m_{\{1,3\}}$}]{};
	\node[fill=gray] (m23) at (3,0) [label=right:{$m_{\{2,3\}}$}]{};
	
	\node[fill=red] (a1) at (-1,2) [label={$a_1$}]{};
	\node[fill=red] (b1) at (1,2) [label={$b_1$}]{};
	\node[fill=green] (a2) at (-3,-2) [label=below:{$a_2$}]{};
	\node[fill=green] (b2) at (-1,-2) [label=below:{$b_2$}]{};
	\node[fill=blue] (a3) at (1,-2) [label=below:{$a_3$}]{};
	\node[fill=blue] (b3) at (3,-2) [label=below:{$b_3$}]{};
	
	\path (a1) edge (m12);
	\path (a1) edge (m13);
	\path (b1) edge (m);
	\path (b1) edge (m23);
	\path (a2) edge (m12);
	\path (a2) edge (m23);
	\path (b2) edge (m);
	\path (b2) edge (m13);
	\path (a3) edge (m23);
	\path (a3) edge (m13);
	\path (b3) edge (m);
	\path (b3) edge (m12);
	\end{tikzpicture}}
	\caption{The CFI-gadget $X_{\{1,2,3\}}$.}\label{cfi:figure}
\end{figure}
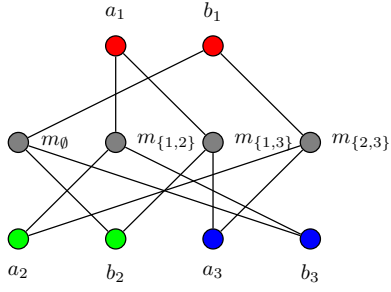

In this section, we look at the graphs by Cai, F{\"u}rer and Immerman \cite{DBLP:journals/combinatorica/CaiFI92}, which were constructed to prove lower bounds for the Weisfeiler-Lehman method in isomorphism testing. These graphs are also challenging when we use resolution to decide isomorphism. They are built from gadget graphs which are defined as follows (see \autoref{cfi:figure}).

\begin{definition}[CFI-gadget {\cite[6]{DBLP:journals/combinatorica/CaiFI92}}] \label{cfi gadget} 
	Given a finite set $N$, define:
		$X_N \coloneqq (V, E, \gamma)$,  where~$V \coloneqq A \cup B \cup M$ consists of~$A \coloneqq \{a_w \mid w\in N \}$,~$B := \{b_w \mid w\in N \}$ as well as~$M \coloneqq \{m_S \mid S \subseteq N, |S| \text{ even} \}$ and~$E := \{\{m_S, a_w\} \mid w\in S \} \cup \{\{m_S, b_w\} \mid w \in N\setminus S\}$.
	Also define the coloring~$\gamma: V \rightarrow C: v \mapsto \gamma(v) := \begin{cases}
			c_w & \text{ if } v \in A\cup B \text{ with } v=a_w \text{ or } v=b_w ,\\
			m & \text{ if } v \in M .\\
			\end{cases}$
\end{definition}

The most important feature of the CFI-gadgets are their automorphisms:

\begin{lemma}\cite[6.1]{DBLP:journals/combinatorica/CaiFI92} \label{cfi symmetry}
	There are $2^{|N|-1}$ automorphisms of $X_N$. Each is uniquely determined by interchanging the vertices $a_w$ and $b_w$ for all $w$ in some subset $S \subseteq N$ of even cardinality.
\end{lemma}

\begin{definition}[CFI graph] \label{cfi graph}
	From a graph $G=(V,E)$ construct $X(G)$ by connecting the CFI gadgets $\{X^v_{\edgeset{G}(v)} \mid v \in V\}$ with edges $E':= \{ \{a^u_e, a^v_e\} \mid e=\{u,v\}\in E \} \cup \{ \{b^u_e, b^v_e\} \mid e=\{u,v\}\in E \}$.
\end{definition}

\begin{definition} \label{tilde_X}
	Given a graph $G=(V,E)$ with $E\neq \emptyset$, construct $\tilde{X}(G)$ from $X(G)$ by choosing some edge $e=\{u,v\}\in E$ and replacing the edges $ \{a^u_e, a^v_e\},\{b^u_e, b^v_e\}$ with the edges~$\{a^u_e, b^v_e\},\{b^u_e, a^v_e\}.$
	We say that the edges corresponding to $e$ have been twisted.
\end{definition}

Note that \autoref{tilde_X} does not specify how to choose the edge which is to be twisted, so there are in fact multiple graphs that we could call $\tilde{X}(G)$. If $G$ is connected however, these graphs are isomorphic. 
On the other hand, for any graph $G$ with at least one edge, $\tilde{X}(G)$ and $X(G)$ are not isomorphic [{\cite[see Lemma 6.2]{DBLP:journals/combinatorica/CaiFI92}}].
The CFI graphs have been used to prove the following lower bound for resolution:

\begin{theorem}[{\cite[Corollary 5.2]{Toran2013}}] 
	\label{thm:torans lower bound}
	There exists a family of graphs $\mathcal{G}=(G_n)_{n\in \mathbb{N}}$ such that for every $n$, $G_n$ has $n$ vertices and the resolution refutation of the formula $F(X(G_n), \tilde{X}(G_n))$ requires
	size $\mathrm{exp}(\Omega(n))$. The graphs $X(G_n)$ and $\tilde{X}(G_n)$ have color multiplicity at most 4.
\end{theorem}

This exponential lower bound motivates the use of a more efficient proof system to prove non-isomorphism of CFI graphs. Because of the symmetric nature of the CFI-gadgets, the symmetry rule is expected to reduce the proof length significantly. With symmetry rule, short proofs exist, as we show in \autoref{sec:cfi res}.

In order to obtain examples for which the symmetry rule is not able to produce short proofs, it is a natural idea to consider asymmetric graphs instead.

\subsection{Multipede Graphs}
In \cite{DBLP:journals/jsyml/GurevichS96}, the so-called Multipedes were defined - a method to construct asymmetric structures. 
Combining this construction with CFI-gadgets, one obtains a family of asymmetric graphs which provide exponential lower bounds for individualization-refinement algorithms \cite{DBLP:conf/stoc/NeuenS18}.

\begin{definition}[Multipede graph \cite{DBLP:conf/stoc/NeuenS18}] \label{multipede graph}
	From a bipartite graph $G=(V,W,E)$, we construct the \emph{Multipede graph} $MP(G)$ as follows: For every $w\in W$ create a pair of vertices $a_w,b_w$, colored with $c_w$. We call these pairs \emph{feet}. Then for every $v \in V$ take a CFI-gadget $X^v_{\neighborhood{G}(v)}$ and identify the vertices $a^v_w$ and $b^v_w$ with $a_w$ and $b_w$ respectively.
\end{definition}

\begin{theorem}[\cite{DBLP:conf/stoc/NeuenS18}]
	There exists a family of bipartite graphs $\mathcal{G}=(G_n)_{n\in \mathbb{N}}$ such that for each $n$ the graph $G_n$ has $\mathcal{O}(n)$ vertices, 
	$MP(G_n)$ is asymmetric and individual\-iza\-tion-refinement algorithms take $\exp(\Omega(n))$ steps to verify $MP(G_n)_{a_\omega} \ncong MP(G_n)_{b_\omega}$.
\end{theorem}

The Multipede graphs are of particular interest, because they are a generalization of the CFI graphs, and thus also hard for resolution, and additionally they can be constructed to be asymmetric. Hence the global symmetry rule is insufficient to get short proofs concerning Multipedes. 
However, as we will prove, the local symmetry of the CFI-gadgets can be used by the local symmetry rule. 

The automorphism group of a Multipede graph is closely related to the solution set of a linear equation system. 
As a consequence of \autoref{cfi symmetry}, any automorphism $\varphi$ of a Multipede can be uniquely specified by the set of feet $Y:= \{w\in W \mid \varphi(a_w) =b_w \}$ for which the $a$-$b$-pairs are swapped. 
The set $Y$ represents a valid automorphism exactly if for every CFI-gadget in the graph, an even number of incident feet is swapped.

Using linear algebra, we can encode a subset $Y\subseteq W=\{w_1, \dots, w_n\}$ uniquely as a vector $\mathbf{y} \in \mathbb{F}_2^n$, by setting $\mathbf{y}_i = 1$ if and only if $w_i \in Y$ for all $i$. 
Then the evenness-condition, which the CFI-gadgets require, can be expressed as a set of linear equations: 
\[ \text{for all $v\in V$: } \sum_{w_i \in N_G(v)} \mathbf{y}_i = 0 \,. \]

We can write the equations in matrix form:
\newcommand{\mpmat}{\mathbf{M}}
	Let $G=(\{v_1,\dots,v_m\}, \{w_1, \dots, w_n\}, E)$ be a bipartite graph.
	Define $\mpmat(G) \in \mathbb{F}_2^{m \times n}$ as follows: 
	$\mpmat(G)_{i,j} := \begin{cases}1 & \text{if } \{v_i, w_j\}\in E\\
	0 & \text{otherwise.}
	\end{cases} $

The solutions of the linear equation system $\mpmat(G) \mathbf{y}=\mathbf{0}$ correspond to the automorphisms of $MP(G)$.
We will show how to apply resolution and the symmetry rule to linear equations, and extend our results to Multipedes.

\subsection{Encoding Linear Equations}
Linear equations over finite fields have been used to show lower bounds in Proof Complexity. For example, the Tseitin formulas are constructed from graphs, representing a system of linear equations over $\mathbb{F}_2$, and are hard for resolution \cite{UrquhartHardRes}.
In the next section we will show that by adding the symmetry rule to resolution, we get short proofs for linear equations.

To work with linear equations, some basic definitions and notations from linear algebra are needed.
Let $K$ be a field and $n,m \in \mathbb{N}$. We write $K^{m \times n}$ for the set of all $m$ by $n$ matrices over $K$. Symbols for matrices will be written in boldface.
Given a matrix $\mathbf{A} \in K^{m \times n}$ and numbers $i\in [1,m]$ and $j\in [1,n]$, we write $\mathbf{A}_{i,j}$ for the element at the $i$-th row and $j$-th column of $\mathbf{A}$.
We write $K^{n}$ for the set of all $n$-element vectors. For our purposes they can be treated like single-column matrices, i.e., $K^n = K^{n \times 1}$.

We write $\mathbf{0}$ for a vector consisting of zeros, where its size is clear from context. Similarly $\mathbf{1}$ is a vector filled with ones.

Applying resolution to a linear equation system means providing a refutation certifying that the system cannot be solved, if that is indeed the case. The following lemma is essential in proving an equation system unsolvable:

\begin{lemma} \label{inconsistent lin eq}
	Let $\mathbf{A} \in \Fp^{m \times n}$ and $\mathbf{b} \in \Fp^m$. If the equation system $\mathbf{A x} = \mathbf{b}$ does not have a solution $\mathbf{x} \in \Fp^n$, then there exists some $\mathbf{v} \in \Fp^m$ such that $\mathbf{v A}=\mathbf{0}$ and $\mathbf{v \cdot b}=1$.
\end{lemma}
\begin{proof}
Since the equation system does not have a solution, applying the Gaussian elimination algorithm yields the equation $0=1$. Writing the row operations used by the algorithm as a vector, we get the sought-after $\mathbf{v}$.
\end{proof}

We require some notation for standard operations from linear algebra.
	Let $\mathbf{r}=(r_1, \dots, r_n) \in K^{n}$ and $\mathbf{A} \in K^{m \times n}$.
		$\supp(\mathbf{r}) := \{i\in\{1,\dots, n\} \mid r_i \neq 0 \}$ is the \emph{support} of $\mathbf{r}$.
		$\mathbf{A}_{i,\ast} := (\mathbf{A}_{i,1},\dots,\mathbf{A}_{i,n}) \in K^{n}$ is the i-th row of $\mathbf{A}$.
		$\mathbf{\diag{r}} \in K^{n \times n}$ is the diagonal matrix with diagonal entries equal to $\mathbf{r}$.
		$\mathbf{\Sigma A} := \sum_{i=1}^m \mathbf{A}_{i,\ast} = \mathbf{1 \cdot A}$ is the row sum of $\mathbf{A}$.
		Let $\mathbf{v}=(v_1, \dots, v_n) \in K^{n}$. 
		Then $\mathbf{r}|_\mathbf{v} \in K^{n}$ is the \emph{restriction} of $\mathbf{r}$ to the support of $\mathbf{v}$, \\defined by $(\mathbf{r}|_\mathbf{v})_i := \begin{cases}
			r_i & \text{if } v_i \neq 0 \\
			0 & \text{if } v_i = 0 \\
		\end{cases}\quad $ for $i\in\{1,\dots, n\}$.

To encode linear equations as CNF formulas, we first introduce variables which correspond to the solution vector of the linear equation system: 
$\mathrm{Vars} := \{ \xi_{i,k} \mid i\in [1,n] \text{ and } k\in \Fp \}$.
For a given vector $\mathbf{x}\in \Fp^n$, the corresponding assignment to the variables would set $\xi_{i,k}$ to \emph{true} if and only if $\mathbf{x}_i=k$.

Our CNF formula has a clause for every $\mathbf{x}$ with $\mathbf{A x} \neq \mathbf{b}$, ensuring the forumla is \emph{false} under the assignment corresponding to $\mathbf{x}$.
For every row $(\mathbf{a},b)$ of the equation system, we consider all $\mathbf{x}$ with $\mathbf{a\cdot x}\neq b$. We can restrict $\mathbf{x}$ to the components for which $\mathbf{a}$ is nonzero. \[P(\mathbf{a},b) := \{\mathbf{x}\in \Fp^n \mid \mathbf{a} \cdot \mathbf{x} \neq b \text{ and } \supp(\mathbf{x}) \subseteq \supp(\mathbf{a})\}.\]
The formula for the row $(\mathbf{a},b)$ is then defined as follows:
\[F(\mathbf{a},b) := \bigwedge_{\mathbf{x}\in P(\mathbf{a}, b)} C_{\mathbf{a}}(\mathbf{x}), \text{ where } C_{\mathbf{a}}(\mathbf{x}) := \bigvee_{i \in \supp(\mathbf{a})} \overline{\xi_{i,\mathbf{x}_i}}.\]
We extend this definition to whole systems of equations: $F(\mathbf{A}, \mathbf{b}):= \bigwedge_{i=1}^{m} F(\mathbf{A}_{i,\ast}, \mathbf{b}_i)$. 
Notice that assigning \emph{false} to every variable satisfies $F(\mathbf{A}, \mathbf{b})$, but this assignment does not represent a vector. For this reason, we additionally need the clauses $V := \bigwedge_{i=1}^{n}\bigvee_{k\in \Fp} \xi_{i,k}$.
\begin{lemma}\label{lem:solvability:vs:lin:alg}
Let $\mathbf{A} \in \Fp^{m \times n}$ and $\mathbf{b} \in \Fp^{m}$. There exists an $\mathbf{x} \in \Fp^{n}$ with $\mathbf{A x = b}$ if and only if $F(\mathbf{A}, \mathbf{b}) \wedge V$ is satisfiable.
\end{lemma}
\begin{proof}
	$\implies$: Assume $\mathbf{A x = b}$. Define an assignment $\varphi: \mathrm{Vars}\rightarrow \binary$ such that $\varphi(\xi_{i,k})=1$ if and only if $\mathbf{x}_i=k$. It is easy to see that $\varphi(V)=1$. Let $j \in [1,m]$ and $\mathbf{x}' \in P(\mathbf{A}_{j,\ast},\mathbf{b}_j)$. Then $\mathbf{A}_{j,\ast} \cdot \mathbf{x}' \neq \mathbf{b}_j = \mathbf{A}_{j,\ast} \cdot \mathbf{x}$. Hence there exists $i\in\supp(\mathbf{A}_{j,\ast})$ such that $\mathbf{x}_i\neq \mathbf{x}'_i$. Therefore $\varphi(\xi_{i, \mathbf{x}'_i}) = 0$, so $\varphi(C_{\mathbf{A}_{j,\ast}}(\mathbf{x}'))=1$. Then $\varphi(F(\mathbf{A}_{j,\ast},\mathbf{b}_j))=1$ and thus $F(\mathbf{A}, \mathbf{b})$ is satisfied by $\varphi$.
	
	$\impliedby$: Assume that we have an assignment $\varphi$ with $\varphi(F(\mathbf{A}, \mathbf{b}))=1$ and $\varphi(V)=1$. For all $i\in[1,n]$ there exists a $k\in\Fp$ such that $\varphi(\xi_{i,k})=1$. Define $\mathbf{x}_i := k$. Towards a contradiction, assume there exists $j \in [1,m]$ with $\mathbf{b}_j \neq \mathbf{A}_{j,\ast} \cdot \mathbf{x}$. Then $\mathbf{x}|_{\mathbf{A}_{j,\ast}} \in P(\mathbf{A}_{j,\ast},\mathbf{b}_j)$ and  $\varphi(C_{\mathbf{A}_{j,\ast}}(\mathbf{x}))=1$. Hence there must exist an $i\in \supp(\mathbf{A}_{j,\ast})$ such that $\varphi(\xi_{i, \mathbf{x}_i})=0$, which contradicts our construction of $\mathbf{x}$. Therefore $\mathbf{A x = b}$.
\end{proof}

\section{Linear-sized Refutations for Non-Isomorphism of CFI graphs}\label{sec:cfi res}
Due to the symmetric nature of the CFI graphs, using the symmetry rule gives us linear-sized resolution proofs of non-isomorphism for a pair of these graphs. 

\begin{theorem}  
	\label{cfi graph symm res}
	Let $G$ be a graph with at least one edge. 
	Then \[ F(X(G),\tilde{X}(G)) \lgsymd_{\mathcal{O}(|F(X(G),\tilde{X}(G))|)} \bot. \]
\end{theorem}
\begin{proof}
	
	Notation: When writing clauses of the isomorphism formula $F(X(G),\tilde{X}(G))$ for the CFI pair, we give different names to the variables $x_{u,v}$, depending on the kind of vertices, to increase readability. This notation is borrowed from \cite{Toran2013} and is as follows:
	\begin{itemize}
		\item \makebox[8cm]{For middle vertices $m^u_S, m^u_{S'}$ we define: \hfill} $z^u_{S,S'} := x_{m^u_S, m^u_{S'}}$
		\item \makebox[8cm]{For a/b vertices $a^u_e, b^u_e$ we define: \hfill} $y_{a^u_e, b^u_e} := x_{a^u_e, b^u_e}$
	\end{itemize}
	
	Definition: For any Graph $G=(V,E)$, we define $\Phi_4(G) := \sum_{v\in V} 4^{\deg_{G}(v)}$. This measure will help to describe the size of the formula $F(X(G),\tilde{X}(G))$.
	For $v\in V$ and $ e\in E$ define $G-e := (V, E\setminus\{e\})$ and $G-v := (V\setminus\{v\}, \{e' \in E \mid v\notin e' \})$. 
	\\\\
	Let $G=(V,E)$ be a graph with $E\neq \emptyset$. We assume that $G$ is connected without loss of generality, otherwise we can apply this proof to the connected component with the twist.
	We will show, for some constant $\alpha$: \[F(X(G),\tilde{X}(G)) \lgsymd_{\alpha |E| + \alpha \Phi_4(G)} \bot \]
	and then $\alpha |E| + \alpha \Phi_4(G) = \mathcal{O}(|F(X(G),\tilde{X}(G))|)$.
	Proceed by induction over the number of edges.\\
	Induction basis: $|E|=1$: 
	The graphs are shown in \autoref{cfi pair 1 edge}.
	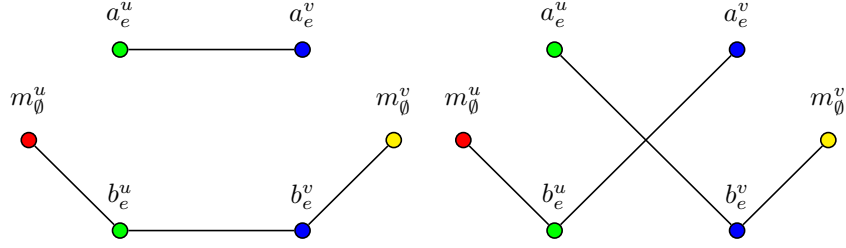
\begin{figure}[h]
		\centering
		\begin{tikzpicture}[scale=1.2,semithick]
		\tikzstyle{every node}=[draw,circle,fill,inner sep=2pt]
		
		\node (mu) at (0,1) [fill=red,label={$m_\emptyset^u$}]{};
		\node (au) at (1,2) [fill=green,label={$a_e^u$}]{};
		\node (bu) at (1,0) [fill=green,label={$b_e^u$}]{};
		\node (av) at (3,2) [fill=blue,label={$a_e^v$}]{};
		\node (bv) at (3,0) [fill=blue,label={$b_e^v$}]{};
		\node (mv) at (4,1) [fill=yellow,label={$m_\emptyset^v$}]{};
		
		\path (mu) edge (bu);
		\path (au) edge (av);
		\path (bu) edge (bv);
		\path (mv) edge (bv);
		\end{tikzpicture}
		\begin{tikzpicture}[scale=1.2,semithick]
		\tikzstyle{every node}=[draw,circle,fill,inner sep=2pt]
		
		\node (mu) at (0,1) [fill=red,label={$m_\emptyset^u$}]{};
		\node (au) at (1,2) [fill=green,label={$a_e^u$}]{};
		\node (bu) at (1,0) [fill=green,label={$b_e^u$}]{};
		\node (av) at (3,2) [fill=blue,label={$a_e^v$}]{};
		\node (bv) at (3,0) [fill=blue,label={$b_e^v$}]{};
		\node (mv) at (4,1) [fill=yellow,label={$m_\emptyset^v$}]{};
		
		\path (mu) edge (bu);
		\path (au) edge (bv);
		\path (bu) edge (av);
		\path (mv) edge (bv);
		\end{tikzpicture}
		\caption{$X(G)$ and $\tilde{X}(G)$ for the connected graph $G$ containing exactly one edge}
		\label{cfi pair 1 edge}
	\end{figure}
	Then $F(X(G),\tilde{X}(G))$ admits the following resolution refutation:
	
	(Notation: ``Type n: C'' means that clause C is contained in $F(X(G),\tilde{X}(G))$. ``$\implies$ C'' signifies that C is resolved from preceding clauses.)
	\begin{align*}
	\text{Type 1: } & z_{\emptyset,\emptyset}^u \\
	\text{Type 3: } & \overline{z_{\emptyset,\emptyset}^u} \vee \overline{y_{b^u_e, a^u_e}} \\
	\implies & \overline{y_{b^u_e, a^u_e}} \\
	\text{Type 1: } & y_{b^u_e, a^u_e} \vee y_{b^u_e, b^u_e} \\
	\implies & y_{b^u_e, b^u_e} \\
	\text{Type 3: } & \overline{y_{b^u_e, b^u_e}} \vee \overline{y_{a^v_e, a^v_e}} \\
	\implies & \overline{y_{a^v_e, a^v_e}} \\
	\text{Type 1: } & y_{a^v_e, a^v_e} \vee y_{a^v_e, b^v_e} \\
	\implies & y_{a^v_e, b^v_e} \\
	\text{Type 3: } & \overline{z_{\emptyset,\emptyset}^v} \vee \overline{y_{a^v_e, b^v_e}} \\
	\implies & \overline{z_{\emptyset,\emptyset}^v} \\
	\text{Type 1: } & z_{\emptyset,\emptyset}^v \\
	\implies & \bot
	\end{align*}
	Here we took $6$ steps to derive $\bot$. We want the following to hold:
	\begin{align*}
	6 &\leq \alpha |E| + \alpha \Phi_4(G) \\
	&= \alpha \cdot 1 + \alpha \cdot (4^1+4^1) = 9 \alpha 
	\end{align*}
	An appropriate $\alpha$ will be determined later. \\\\
	Induction step: 
	\\
	Case 1: $G$ is a tree. \\
	Pick a vertex $u$ of $G$ with exactly one incident edge $e=\{u,v\}$, where the corresponding edges in $\tilde{X}(G)$ are not twisted. This is possible since $G$ is a tree by assumption, so it has at least two vertices of degree 1. Note that $X(G)$ and $\tilde{X}(G)$ are locally identical in this case. The situation is shown in \autoref{fig 2}. 
	Define $\mathcal{S}^v := \{ A \subseteq \edgeset{G}(v) \mid |A| \text{ even}\}$. 
	We have $|\mathcal{S}^v| = 2^{|\edgeset{G}(v)|-1} = 2^{\deg_G(v)-1}$.
	Then resolve the following clauses:
	\begin{align*}
	&\text{Type 1: } && \hfill z_{\emptyset,\emptyset}^u \\
	&\text{Type 3: } && \overline{z_{\emptyset,\emptyset}^u} \vee \overline{y_{a_e^u,b_e^u}} \\
	&\implies && \overline{y_{a_e^u,b_e^u}} \\
	&\text{Type 1: } && y_{a_e^u,a_e^u} \vee y_{a_e^u,b_e^u} \\
	&\implies && y_{a_e^u,a_e^u} \\
	&\text{Type 3: } && \overline{y_{a_e^u,a_e^u}} \vee \overline{y_{a_e^v,b_e^v}} \\
	&\implies && \overline{y_{a_e^v,b_e^v}} \\
	&\text{Type 1: } && y_{a_e^v,a_e^v} \vee y_{a_e^v,b_e^v} \\
	&\implies && y_{a_e^v,a_e^v} \\
	\forall S,S'\in \mathcal{S}^v \text{ with } e \notin S, e \in S': &\text{Type 3: } && \overline{z_{S,S'}^v} \vee \overline{y_{a_e^v,a_e^v}} \\
	\forall S,S'\in \mathcal{S}^v \text{ with } e \notin S, e \in S': &\implies && \overline{z_{S,S'}^v} \\
	\forall S\in \mathcal{S}^v, e \notin S: &\text{Type 1: } && \bigvee_{S' \in \mathcal{S}^v} z_{S,S'}^v \\
	&& & \equiv \bigvee_{S' \in \mathcal{S}^v, e \notin S'} z_{S,S'}^v \vee \bigvee_{S' \in \mathcal{S}^v, e \in S'} z_{S,S'}^v\\
	\forall S\in \mathcal{S}^v, e \notin S: &\implies && \bigvee_{S' \in \mathcal{S}^v, e \notin S'} z_{S,S'}^v 
	\end{align*}
	This resolution takes $4 + \tfrac{1}{4}|\mathcal{S}^v|^2 + \tfrac{1}{4}|\mathcal{S}^v|^2$ steps. 
	We obtain the Type 1 clauses of $F(X(G-u),\tilde{X}(G-u))$. The Type 2 and 3 clauses of $F(X(G-u),\tilde{X}(G-u))$ are already present in $F(X(G),\tilde{X}(G))$. 
	
	By induction, $F(X(G-u),\tilde{X}(G-u))$ can be resolved to $\bot$ in $\alpha |E\setminus \{e\}| + \alpha \Phi_4(G-u)$ steps.
	It holds:
	\begin{align*}
	\Phi_4(G-u) &= \sum_{\mu \in V\setminus\{u\}} 4^{\deg_{G-u}(\mu)}\\
	&= 4^{\deg_{G-u}(v)} + \sum_{\mu \in V\setminus\{u,v\}} 4^{\deg_{G-u}(\mu)}\\
	&= 4^{\deg_{G}(v)-1} + \sum_{\mu \in V\setminus\{u,v\}} 4^{\deg_{G}(\mu)}\\
	&= \tfrac{1}{4} 4^{\deg_{G}(v)} + \Phi_4(G) - 4^{\deg_{G}(u)} - 4^{\deg_{G}(v)}\\
	&= \Phi_4(G) - 4 - \tfrac{3}{4} 4^{\deg_{G}(v)} .
	\end{align*}
	We calculate the total number of resolution steps $T$ for this case:
	\begin{align*}
	\implies T &= 4 + \tfrac{1}{4}|\mathcal{S}^v|^2 + \tfrac{1}{4}|\mathcal{S}^v|^2 + \alpha |E\setminus \{e\}| + \alpha \Phi_4(G-u) \\
	&= 4 + \tfrac{1}{2}(2^{\deg_G(v)-1})^2 + \alpha |E| - \alpha + \alpha \Phi_4(G-u) \\
	&= 4 + \tfrac{1}{8}4^{\deg_G(v)} + \alpha |E| - \alpha + \alpha (\Phi_4(G) - 4 - \tfrac{3}{4} 4^{\deg_{G}(v)})
	\end{align*}
	Then our bound for $T$ must hold:
	\begin{align*}
	&&T &\leq \alpha |E| + \alpha \Phi_4(G) \\
	&\iff& 4 + \tfrac{1}{8}4^{\deg_G(v)} + \alpha |E| - \alpha + \alpha (\Phi_4(G) - 4 - \tfrac{3}{4} 4^{\deg_{G}(v)}) &\leq \alpha |E| + \alpha \Phi_4(G) \\
	&\iff& 4 + \tfrac{1}{8}4^{\deg_G(v)} - \alpha + \alpha ( - 4 - \tfrac{3}{4} 4^{\deg_{G}(v)}) &\leq 0 \\
	&\iff& 4 + \tfrac{1}{8}4^{\deg_G(v)} &\leq 5 \alpha + \tfrac{3}{4} \alpha 4^{\deg_{G}(v)}
	\end{align*}
	We can satisfy this by requiring
	\begin{align*}
	4 \leq 5 \alpha \text{ and } \tfrac{1}{8}\leq \tfrac{3}{4} \alpha .
	\end{align*}
	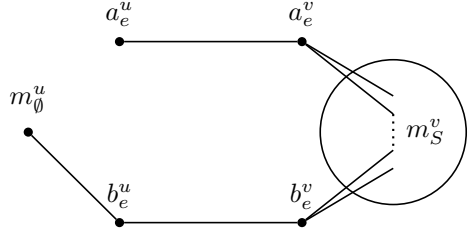
\begin{figure}
		\centering
		\begin{tikzpicture}[scale=1.2,semithick]
		\tikzstyle{every node}=[draw,circle,fill,inner sep=1pt]
		
		\node (mu) at (0,1) [label={$m_\emptyset^u$}]{};
		\node (au) at (1,2) [label={$a_e^u$}]{};
		\node (bu) at (1,0) [label={$b_e^u$}]{};
		\node (av) at (3,2) [label={$a_e^v$}]{};
		\node (bv) at (3,0) [label={$b_e^v$}]{};
		\node (mv) at (4,1) [label=right:{$m_S^v$},draw=none,fill=none]{};
		
		\path (mu) edge (bu);
		\path (au) edge (av);
		\path (bu) edge (bv);
		
		\draw (4,1) circle (0.8);
		\draw (av) -- (4,1+0.2);
		\draw (av) -- (4,1+0.4);
		\draw (bv) -- (4,1-0.2);
		\draw (bv) -- (4,1-0.4);
		\draw[thick,dotted] (4,1+0.2) -- (4,1-0.2);
		\end{tikzpicture}
		\caption{Vertex of degree 1 in $X(G)$} \label{fig 2}
	\end{figure}
	\\\\
	Case 2: $G$ has a cycle.\\
	We choose a simple cycle, i.e.\ one which does not repeating vertices.
	Pick an arbitrary edge $e=\{u,v\}$ along the cycle, such that the corresponding edge in $\tilde{X}(G)$ is not twisted (see \autoref{cfi higher deg}), and resolve the clauses:
	\begin{align*}
	&\text{Type 3: } && \overline{y_{a_e^u,a_e^u}} \vee \overline{y_{a_e^v,b_e^v}} \\
	&\text{Type 1: } && y_{a_e^v,a_e^v} \vee y_{a_e^v,b_e^v} \\
	&\implies && \overline{y_{a_e^u,a_e^u}} \vee y_{a_e^v,a_e^v} \label{c2eq11}\\
	\forall \mu\in e: \forall S,S'\in \mathcal{S}^\mu, e \notin S, e \in S': &\text{Type 3: } && \overline{z_{S,S'}^\mu} \vee \overline{y_{a_e^\mu,a_e^\mu}} \\
	\forall \mu\in e: \forall S,S'\in \mathcal{S}^\mu, e \notin S, e \in S': &\implies && \overline{z_{S,S'}^\mu} \vee \overline{y_{a_e^u,a_e^u}} \\
	\forall \mu\in e: \forall S\in \mathcal{S}^\mu, e \notin S: &\text{Type 1: } && \bigvee_{S' \in \mathcal{S}^\mu} z_{S,S'}^\mu \\
	\forall \mu\in e: \forall S\in \mathcal{S}^\mu, e \notin S: &\implies && \overline{y_{a_e^u,a_e^u}} \vee \bigvee_{S' \in \mathcal{S}^\mu, e\notin S'} z_{S,S'}^\mu 
	\end{align*}
	We obtain the Type 1 clauses of $F(X(G-e),\tilde{X}(G-e))$, but with some extra literals $\overline{y_{a_e^u,a_e^u}}$.
	Since $e$ is part of a cycle, $G-e$ is still connected. By induction, $F(X(G-e),\tilde{X}(G-e))$ can be resolved to $\bot$; 
	consequently, $F(X(G),\tilde{X}(G))$ can be resolved to $\overline{y_{a_e^u,a_e^u}}$.
	Together, this takes \[
	1 + \tfrac{1}{4}|\mathcal{S}^v|^2 + (\tfrac{1}{4}|\mathcal{S}^u|^2+\tfrac{1}{4}|\mathcal{S}^v|^2) + \alpha |E\setminus \{e\}| + \alpha \Phi_4(G-e)
	\] resolution steps.
	
	By deriving $\overline{y_{a_e^u,a_e^u}}$ we showed that there is no isomorphism from $X(G)$ to $\tilde{X}(G)$ which maps $a_e^u$ to $a_e^u$. Hence the only way an isomorphism could exist is by mapping $b_e^u$ to $a_e^u$. But we will see that $X(G)$ has an automorphism swapping $a_e^u$ with $b_e^u$, so we can apply the same reasoning to show that an isomorphism also cannot map $b_e^u$ to $a_e^u$.
	
	This is how we exploit the symmetries of the CFI graphs: Let $C$ be the set of vertices of $G$ that lie on the cycle. Then for every $x\in C$, exactly two of its neighbors are in $C$. By \autoref{cfi symmetry}, the CFI-gadget corresponding to $x$ has an automorphism exchanging the two $a$-$b$-pairs related to these neighbors. We assemble these automorphisms together for every $x\in C$ to obtain an automorphism $\varphi$ of $X(G)$ that swaps all $a$-$b$-pairs along the cycle.
	
	Then $\varphi(a_e^u)=b_e^u$. 
	This automorphism of the graph induces a symmetry $\psi$ on the formula $F(X(G),\tilde{X}(G))$ 
	with $\psi(y_{a_e^u,a_e^u})=y_{b_e^u,a_e^u}$. Apply the global symmetry rule
	on the resolution of $\overline{y_{a_e^u,a_e^u}}$ to obtain $\overline{y_{b_e^u,a_e^u}}$. Resolve:
	\begin{align*}
	\text{Previously: } && \overline{y_{a_e^u,a_e^u}} \\
	\text{Symmetry rule} \implies && \overline{y_{b_e^u,a_e^u}} \\
	\text{Type 1: } && y_{a_e^u,a_e^u} \vee y_{a_e^u,b_e^u} \\
	\text{Type 2: } && \overline{y_{a_e^u,b_e^u}} \vee \overline{y_{b_e^u,b_e^u}} \\
	\implies && y_{a_e^u,a_e^u} \vee \overline{y_{b_e^u,b_e^u}}\\
	\text{Type 1: } && y_{b_e^u,a_e^u} \vee y_{b_e^u,b_e^u} \\
	\implies && y_{b_e^u,a_e^u} \vee y_{a_e^u,a_e^u}\\
	\implies && y_{b_e^u,a_e^u} \\
	\implies && \bot
	\end{align*}
	Then the total number of steps is
	\begin{align*}
	T &= 1 + \tfrac{1}{4}|\mathcal{S}^v|^2 + (\tfrac{1}{4}|\mathcal{S}^u|^2+\tfrac{1}{4}|\mathcal{S}^v|^2) + \alpha |E\setminus \{e\}| + \alpha \Phi_4(G-e) + 5 \\
	&= 6 + \tfrac{1}{8}4^{\deg_{G}(v)} + \tfrac{1}{16}4^{\deg_{G}(u)} + \alpha |E| - \alpha + \alpha \Phi_4(G-e) ,\\
	\text{with }\;\; \Phi_4(G-e) &= \sum_{\mu \in V} 4^{\deg_{G-e}(\mu)}\\
	&= \sum_{\mu \in V\setminus\{u,v\}} 4^{\deg_{G-e}(\mu)} + 4^{\deg_{G-e}(u)} + 4^{\deg_{G-e}(v)} \\
	&= \sum_{\mu \in V\setminus\{u,v\}} 4^{\deg_{G}(\mu)} + 4^{\deg_{G}(u)-1} + 4^{\deg_{G}(v)-1} \\
	&= \Phi_4(G) - \tfrac{3}{4} 4^{\deg_{G}(u)} - \tfrac{3}{4} 4^{\deg_{G}(v)} .
	\end{align*}
	Then
	\begin{align*}
	&&T \leq \alpha |E| + \alpha \Phi_4(G) \\
	&\iff& 6 + \tfrac{1}{8}4^{\deg_{G}(v)} + \tfrac{1}{16}4^{\deg_{G}(u)} - \alpha + \alpha ( - \tfrac{3}{4} 4^{\deg_{G}(u)} - \tfrac{3}{4} 4^{\deg_{G}(v)}) \leq 0  \\
	&\iff& 6 + \tfrac{1}{8}4^{\deg_{G}(v)} + \tfrac{1}{16}4^{\deg_{G}(u)} \leq \alpha + \alpha ( \tfrac{3}{4} 4^{\deg_{G}(u)} + \tfrac{3}{4} 4^{\deg_{G}(v)})
	\end{align*}
	This is satisfied by the constraint
	\begin{align*}
	6 \leq \alpha \wedge \tfrac{1}{8} \leq \tfrac{3}{4} \alpha \wedge \tfrac{1}{16} \leq \tfrac{3}{4} \alpha .
	\end{align*}
	\begin{figure}
		\centering
		\begin{tikzpicture}[scale=1.2,semithick]
		\tikzstyle{every node}=[draw,circle,fill,inner sep=1pt]
		
		\node (mu) at (0,1) [label=left:{$m_S^u$},draw=none,fill=none]{};
		\node (au) at (1,2) [label={$a_e^u$}]{};
		\node (bu) at (1,0) [label={$b_e^u$}]{};
		\node (av) at (3,2) [label={$a_e^v$}]{};
		\node (bv) at (3,0) [label={$b_e^v$}]{};
		\node (mv) at (4,1) [label=right:{$m_S^v$},draw=none,fill=none]{};
		
		\path (au) edge (av);
		\path (bu) edge (bv);
		
		\draw (4,1) circle (0.8);
		\draw (av) -- (4,1+0.2);
		\draw (av) -- (4,1+0.4);
		\draw (bv) -- (4,1-0.2);
		\draw (bv) -- (4,1-0.4);
		\draw[thick,dotted] (4,1+0.2) -- (4,1-0.2);
		\draw (0,1) circle (0.8);
		\draw (au) -- (0,1+0.2);
		\draw (au) -- (0,1+0.4);
		\draw (bu) -- (0,1-0.2);
		\draw (bu) -- (0,1-0.4);
		\draw[thick,dotted] (0,1+0.2) -- (0,1-0.2);
		\end{tikzpicture}
		\caption{Vertices of higher degree in $X(G)$}
		\label{cfi higher deg}
	\end{figure}
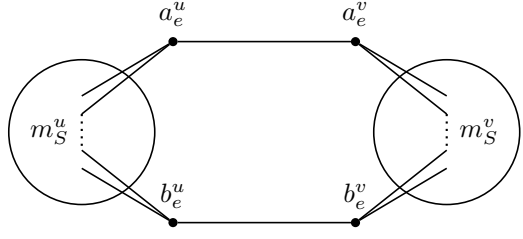
	Using $\alpha=6$ will suffice. Lastly, we need to show that $\alpha |E| + \alpha \Phi_4(G)$ is linearly bounded by the size of the formula $F(X(G),\tilde{X}(G))$. We have:
	\begin{align*}
	|E| = \tfrac{1}{2} \sum_{v\in V} \deg_{G}(v) \leq \tfrac{1}{2} \sum_{v\in V} 4^{\deg_{G}(v)} = \mathcal{O}(\Phi_4(G))
	\end{align*}
	Now count the clauses of $F(X(G),\tilde{X}(G))$: For every $v \in V$ we have $2^{\deg_{G}(v)-1}$ middle vertices in both $X(G)$ and $\tilde{X}(G)$. This results in \[\sum_{v\in V} \begin{pmatrix}
	2^{\deg_{G}(v)-1} \\ 
	2
	\end{pmatrix} = \sum_{v\in V} \tfrac{1}{2} 2^{\deg_{G}(v)-1}(2^{\deg_{G}(v)-1}-1) \] clauses of Type 2 and
	\[
	\sum_{v\in V} 2^{\deg_{G}(v)-1}
	\] clauses of Type 1. Thus
	\begin{align*}
	|F(X(G),\tilde{X}(G))| &\geq \sum_{v\in V} \tfrac{1}{2} 2^{\deg_{G}(v)-1}(2^{\deg_{G}(v)-1}-1) + \sum_{v\in V} 2^{\deg_{G}(v)-1} \\
	&= \sum_{v\in V} \tfrac{1}{2} 2^{\deg_{G}(v)-1}2^{\deg_{G}(v)-1} - \sum_{v\in V} \tfrac{1}{2} 2^{\deg_{G}(v)-1} + \sum_{v\in V} 2^{\deg_{G}(v)-1} \\
	&= \tfrac{1}{2} \sum_{v\in V} 4^{\deg_{G}(v)-1} + \tfrac{1}{2} \sum_{v\in V} 2^{\deg_{G}(v)-1} \\
	&\geq \tfrac{1}{8} \Phi_4(G) 
	\end{align*}
	\begin{align*}
	\implies \alpha |E| + \alpha \Phi_4(G) = \mathcal{O}(\Phi_4(G)) = \mathcal{O}(|F(X(G),\tilde{X}(G))|) \qquad \qedhere
	\end{align*}
\end{proof}

This result stands in contrast to the exponential lower bound of \autoref{thm:torans lower bound}.

\section{Polynomial-sized Refutations for Linear Equations}\label{sec:lin:eqs}
\subsection{Linear Combinations}
The usual approach to showing that a system of linear equations is inconsistent, is to build a linear combination of the equations to derive the obvious contradiction $0=1$. This method is complete by \autoref{inconsistent lin eq}.
We recreate this process in the resolution proof system. However we need to ensure that the support of equations we create along the way is not excessively large. We will do so by using the the symmetry rule.

Note that the formula $F(\mathbf{a},b)$ is invariant under linear scaling of the inputs: For any $k \in \Fp\setminus\{0\}$ we have $\supp(\mathbf{a})=\supp(k\mathbf{a})$ and $P(\mathbf{a},b) = P(k\mathbf{a},k b)$.
Hence $F(\mathbf{a}, b) = F(k\mathbf{a}, k b)$.

For the computation of linear combinations, we use the following definition.
For $\theta \subseteq [1,n]$ define $\Omega(\theta) := \left\{\bigvee_{i \in \theta} \overline{\xi_{i,\mathbf{x}_i}} \mid \mathbf{x}\in \Fp^n \text{ with } \supp(\mathbf{x}) \subseteq \theta \right\}$. Together, the clauses in $\Omega(\theta)$ forbid all possible assignments to the components in range $\theta$. In a sense, $\Omega(\theta)$ is our basic building block for contradictions.
\begin{lemma}\label{thmOmega}
	Let $\theta \subseteq [1,n]$.
	Then $\Omega(\theta) \wedge V \vdash_{\frac{p^{|\theta|+1}-p}{p-1}} \bot$.
\end{lemma}
\begin{proof}
	Induction over $|\theta|$. If $\theta=\emptyset$ then $\Omega(\theta)=\{ \bigvee \emptyset \} = \{\bot\}$, so $\Omega(\theta) \vdash_{0} \bot$.
	
	Induction step: $\theta = \theta' \cup \{j\}$. 
	It holds: 
	\begin{align*}
	\Omega(\theta) &= \left\{\bigvee_{i \in \theta} \overline{\xi_{i,\mathbf{x}_i}} \mid \mathbf{x}\in \Fp^n \text{ with } \supp(\mathbf{x}) \subseteq \theta \right\} \\
	&= \left\{\overline{\xi_{j,\mathbf{x}_j}} \vee \bigvee_{i \in \theta'} \overline{\xi_{i,\mathbf{x}_i}} \mid \mathbf{x}\in \Fp^n \text{ with } \supp(\mathbf{x}) \subseteq \theta \right\} \\
	&= \left\{\overline{\xi_{j,k}} \vee \bigvee_{i \in \theta'} \overline{\xi_{i,\mathbf{x}_i}} \mid \mathbf{x}\in \Fp^n \text{ with } \supp(\mathbf{x}) \subseteq \theta' \text{ and } k\in \Fp \right\}\\
	&= \left\{\overline{\xi_{j,k}} \vee c' \mid c'\in \Omega(\theta') ,  k \in \Fp \right\}  
	\end{align*}
	For each $c'\in \Omega(\theta')$, we can derive the clause $c'$ by resolving $\bigvee_{k\in \Fp} \xi_{j,k}$ from $V$ with the clauses from $\Omega(\theta)$. Doing this for all $c'\in \Omega(\theta')$ takes $p \cdot |\Omega(\theta')| = |\Omega(\theta)| = p^{|\theta|}$ resolution steps.
	By induction, we can then derive $\bot$ from $\Omega(\theta')$ and $V$ in $\frac{p^{|\theta'|+1}-p}{p-1}=\frac{p^{|\theta|}-p}{p-1}$ steps. The total number of steps taken is $\frac{p^{|\theta|}-p}{p-1} + p^{|\theta|} = \frac{p^{|\theta|+1}-p}{p-1}$.
\end{proof}

\begin{figure}
\begin{center}
\begin{tikzpicture}[scale=.35]
\filldraw[thin,red,opacity=.3] (0, 0) rectangle (7,1);
\node at (-1.7,0.5) {$\mathbf{x}$};

\filldraw[thin,red,opacity=.3] (3, 0-1.1) rectangle (10, 1-1.1);
\node at (-1.7, 0.5-1.1) {$\mathbf{y}$};

\filldraw[thin,red,opacity=.3] (0, 0-2*1.1) rectangle (4, 1-2*1.1);
\filldraw[thin,red,opacity=.3] (5, 0-2*1.1) rectangle (6, 1-2*1.1);
\filldraw[thin,red,opacity=.3] (7, 0-2*1.1) rectangle (10, 1-2*1.1);
\node at (-1.7, 0.5-2*1.1) {$\mathbf{x+y}$};

\filldraw[thin,red,opacity=.3] (4, 0-3*1.1) rectangle (5, 1-3*1.1);
\filldraw[thin,red,opacity=.3] (6, 0-3*1.1) rectangle (7, 1-3*1.1);
\node at (-1.7, 0.5-3*1.1) {$\lv(\mathbf{x}, \mathbf{y})$};
\end{tikzpicture}
\end{center}
\caption{A visualization of $\lv(\mathbf{x}, \mathbf{y})$.}\label{figure:for:sum:vanish}
\end{figure}
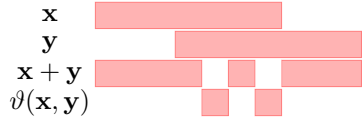

When we sum two vectors $\mathbf{x}$ and $\mathbf{y}$, some components may become zero which were nonzero before. The following definition captures this phenomenon: $\lv(\mathbf{x}, \mathbf{y}) := (\supp(\mathbf{x}) \cup \supp(\mathbf{y})) \setminus \supp(\mathbf{x+y})$.
If a coefficient vanishes in a sum, it has to appear in both summands: $\lv(\mathbf{x}, \mathbf{y}) \subseteq \supp(\mathbf{x}) \cap \supp(\mathbf{y})$ (see \autoref{figure:for:sum:vanish}). 

With these ingredients, we can finally explain the process of building sums using resolution.

\begin{theorem}[Sum Resolution] \label{sum resolution}
	Let $\mathbf{a} \in \Fp^{2\times n}$ and $\mathbf{b} \in \Fp^2$.
	Define $\theta := \lv(\mathbf{a}_1, \mathbf{a}_2)$, where~$\mathbf{a}_i$ is the~$i$-th row of~$a$.
	For all $c\in F(\mathbf{\Sigma a}, \mathbf{\Sigma b})$ it holds:
	$F(\mathbf{a}_1, \mathbf{b}_1) \cup F(\mathbf{a}_2, \mathbf{b}_2) \cup V \vdash^w_{2(p^{|\theta|}-1)} c$. 
\end{theorem}
\begin{proof}
	Let $c \in F(\mathbf{\Sigma a}, \mathbf{\Sigma b})$. By definition of $F$, there exists some $\mathbf{x}\in P(\mathbf{\Sigma a}, \mathbf{\Sigma b})$ such that $c= \bigvee_{i \in \supp(\mathbf{\Sigma a})} \overline{\xi_{i,\mathbf{x}_i}}$.
	The resolution derivation will have to get rid of all variables corresponding to components in $\theta$.
	For $\kappa\in\{1,2\}$ define $R^\kappa := \bigvee_{i \in \supp(\mathbf{a}^\kappa) \setminus \theta} \overline{\xi_{i, \mathbf{x}_i}}$.
	From this we will build the desired clause $c$. It holds: $(\supp(\mathbf{a}_1) \cup \supp(\mathbf{a}_2)) \setminus \theta = (\supp(\mathbf{a}_1) \cup \supp(\mathbf{a}_2)) \cap \supp{(\mathbf{\Sigma a})} = \supp{(\mathbf{\Sigma a})}$.
	Thus
	\begin{align*}
	R^1 \vee R^2 &= \bigvee_{i \in \supp(\mathbf{a}_1) \setminus \theta} \overline{\xi_{i, \mathbf{x}_i}} \vee \bigvee_{i \in \supp(\mathbf{a}_2) \setminus \theta} \overline{\xi_{i, \mathbf{x}_i}} \\
	&= \bigvee_{i \in (\supp(\mathbf{a}_1) \cup \supp(\mathbf{a}_2)) \setminus \theta} \overline{\xi_{i, \mathbf{x}_i}} \\
	&= \bigvee_{i \in \supp(\mathbf{\Sigma a})} \overline{\xi_{i, \mathbf{x}_i}} = c
	\end{align*}
	
	Consider an arbitrary $\mathbf{y}\in \Fp^n$ with $\supp(\mathbf{y}) \subseteq \theta$. Since $\supp(\mathbf{y}) \cap \supp(\mathbf{\Sigma a})=\emptyset$, we have $\mathbf{\Sigma a} \cdot \mathbf{y}=0$. There exists $\kappa$ with $\mathbf{a}_\kappa \cdot (\mathbf{x+y}) \neq b_\kappa$, because otherwise we would have $\mathbf{\Sigma b} = \mathbf{b}_1+\mathbf{b}_2 = \mathbf{a}_1 \cdot (\mathbf{x+y}) + \mathbf{a}_2 \cdot (\mathbf{x+y}) = (\mathbf{a}_1+\mathbf{a}_2) \cdot (\mathbf{x+y}) = \mathbf{\Sigma a} \cdot \mathbf{x} + \mathbf{\Sigma a} \cdot \mathbf{y}$ = $\mathbf{\Sigma a}\cdot \mathbf{x}$, which contradicts $\mathbf{x}\in P(\mathbf{\Sigma a}, \mathbf{\Sigma b})$.
	
	Hence $(\mathbf{x}+\mathbf{y})|_{\mathbf{a}_\kappa} \in P(\mathbf{a}_\kappa, \mathbf{b}_\kappa)$
	and we have a clause $c'(\mathbf{y}) := \bigvee_{i \in \supp(\mathbf{a}_\kappa)} \overline{\xi_{i,\mathbf{(x+y)}_i}} \in F(\mathbf{a}_\kappa, \mathbf{b}_\kappa)$.
	It holds:
	\begin{alignat*}{3}
	c'(\mathbf{y}) &= \bigvee_{i \in \supp(\mathbf{a}_\kappa)\cap \theta} &\overline{\xi_{i,\mathbf{(x+y)}_i}} &\vee \bigvee_{i \in \supp(\mathbf{a}_\kappa) \setminus \theta} &\overline{\xi_{i,\mathbf{(x+y)}_i}} \\
	&= \bigvee_{i \in \supp(\mathbf{a}_\kappa)\cap \theta} &\overline{\xi_{i,\mathbf{y}_i}} &\vee \bigvee_{i \in \supp(\mathbf{a}_\kappa) \setminus \theta} &\overline{\xi_{i,\mathbf{x}_i}} \\
	&= \bigvee_{i \in \theta} &\overline{\xi_{i,\mathbf{y}_i}} &\vee &R^\kappa 
	\end{alignat*}
	
	Now, looking at the set $C:=\{c'(\mathbf{y}) \mid \mathbf{y}\in \Fp^n\}\subseteq F(\mathbf{a}_1, \mathbf{b}_1) \cup F(\mathbf{a}_2, \mathbf{b}_2)$, note that for every clause $d\in \Omega(\theta)$, we have $d \vee R^1 \in C$ or $d \vee R^2 \in C$. By \autoref{thmOmega} and \autoref{reshelper2} 
	we can resolve the clauses in $C$ together with $V$ to obtain $R^1 \vee R^2 = c$ or stronger. 
	Since $p\geq2$, this takes at most $\frac{p^{|\theta|+1}-p}{p-1} \leq \frac{p^{|\theta|+1}-p}{p/2} = 2(p^{|\theta|}-1)$ resolution steps. 
\end{proof}

By applying \autoref{sum resolution} iteratively, we can construct the formulas for linear combinations with an arbitrary number of summands. This method, however, is inefficient since the produced intermediate equations may accumulate more and more variables, leading to an exponential growth of the number of required clauses.

We solve this problem by deriving only a single representative clause for intermediate results, and using the local symmetry rule to derive more clauses as necessary.

\subsection{Local Symmetry in Equations}
We want to understand which symmetries the formulas corresponding to linear equations have. For $\mathbf{d} \in \Fp^n$ define $\Delta_{\mathbf{d}}: \mathrm{Vars}\rightarrow
\mathrm{Vars}: \xi_{i,k} \mapsto \Delta_{\mathbf{d}}(\xi_{i,k}) := \xi_{i,k+\mathbf{d}_i}$. This bijective map is a translation by $\mathbf{d}$ of the vector corresponding to the variables. 

\begin{lemma} \label{Delta}
	Let $b \in \Fp$ and $\mathbf{a},\mathbf{d} \in \Fp^n$. 
	Then $\Delta_\mathbf{d} \in \mathrm{Sym}(F(\mathbf{a},b))$ if and only if
	$\mathbf{a} \cdot \mathbf{d} = 0$. 
\end{lemma}

\begin{proof}
	$\impliedby$: Assume $\mathbf{a} \cdot \mathbf{d} = 0$. Let $c = C_{\mathbf{a}}(\mathbf{x}) = \bigvee_{i \in \supp(\mathbf{a})} \overline{\xi_{i,\mathbf{x}_i}} \in F(\mathbf{a},b)$ for some $\mathbf{x} \in P(\mathbf{a},b)$.
	We have $\mathbf{a} \cdot (\mathbf{x+d}) = \mathbf{a} \cdot \mathbf{x} \neq b$. Hence $\mathbf{(x+d)}|_{\mathbf{a}} \in P(\mathbf{a},b)$ and thus $\Delta_\mathbf{d}(c) = \bigvee_{i \in \supp(\mathbf{a})} \overline{\xi_{i,\mathbf{x}_i+\mathbf{d}_i}} = C_{\mathbf{a}}(\mathbf{x+d}) = C_{\mathbf{a}}(\mathbf{(x+d)}|_{\mathbf{a}}) \in F(\mathbf{a},b)$.
	
	$\implies$: Assume $\mathbf{a} \cdot \mathbf{d} \neq 0$. Then $\mathbf{a} \neq \mathbf{0}$ and hence there exists a vector $\mathbf{x} \in P(\mathbf{a},b)$ with $\mathbf{a \cdot x}= b - \mathbf{a} \cdot \mathbf{d} \neq b$. But then $\mathbf{a} \cdot (\mathbf{x+d})=b$ and thus $\Delta_\mathbf{d}(C_{\mathbf{a}}(\mathbf{x})) = C_{\mathbf{a}}(\mathbf{x+d}) \notin F(\mathbf{a},b)$. Hence $\Delta_\mathbf{d} \notin \mathrm{Sym}(F(\mathbf{a},b))$.
\end{proof}
\begin{corollary}\label{cor:sym:to:lin:alg}
	Let $\mathbf{A} \in \Fp^{m \times n}$, $\mathbf{b} \in \Fp^{m}$ and $\mathbf{d} \in \Fp^{n}$.
	If $\mathbf{A} \cdot \mathbf{d} = \mathbf{0}$, then $\Delta_\mathbf{d} \in \mathrm{Sym}(F(\mathbf{A},\mathbf{b}))$.
\end{corollary}

Note the following: If $\mathbf{d},\mathbf{d'} \in \Fp^n$ such that $\mathbf{d}|_{\mathbf{a}}=\mathbf{d'}|_{\mathbf{a}}$, then for all $c \in F(\mathbf{a},b)$ it holds: $\Delta_{\mathbf{d}}(c) = \Delta_{\mathbf{d'}}(c)$. In particular, $\Delta_\mathbf{d} \in \mathrm{Sym}(F(\mathbf{a},b))$ implies $\Delta_\mathbf{d'} \in \mathrm{Sym}(F(\mathbf{a},b))$.
The condition $\mathbf{d}|_{\mathbf{a}}=\mathbf{d'}|_{\mathbf{a}}$ can equivalently be expressed using matrix algebra: $\mathbf{\diag{a} d} = \mathbf{\diag{a} d'}$.

To make use of the symmetry rule, we want to apply the symmetries of $F(\mathbf{A},\mathbf{b})$ to derive clauses of $F(\mathbf{\Sigma A},\mathbf{\Sigma b})$. From the statements and \autoref{cor:sym:to:lin:alg}, we conclude the following relation between $\mathrm{Sym}(F(\mathbf{A}, \mathbf{b}))$ and $\mathrm{Sym}(F(\mathbf{\Sigma A}, \mathbf{\Sigma b}))$.
\begin{lemma} \label{symsummatrix}
	Let $\mathbf{A} \in \Fp^{m \times n}$, $\mathbf{b} \in \Fp^{m}$ and $\mathbf{d} \in \Fp^{n}$ with $\Delta_\mathbf{d} \in \mathrm{Sym}(F(\mathbf{\Sigma A}, \mathbf{\Sigma b}))$ 
	If there exists $\mathbf{d'} \in \Fp^{n}$ such that $\mathbf{A d'} = \mathbf{0}$ and $\mathbf{\diag{\Sigma A} d'} = \mathbf{\diag{\Sigma A} d}$, then
	$\Delta_{\mathbf{d'}} \in \mathrm{Sym}(F(\mathbf{A}, \mathbf{b}))$ and 
	$\Delta_{\mathbf{d'}}(c) = \Delta_{\mathbf{d}}(c)$ for all $c\in F(\mathbf{\Sigma A}, \mathbf{\Sigma b})$.
\end{lemma}

Concerning $V$, the symmetries are simpler: For any $\mathbf{d} \in \Fp^{n}$ we have $\Delta_{\mathbf{d}} \in \mathrm{Sym}(V)$.

We will assume that the coefficient matrices $\mathbf{A}$ in the following have at most $L$ nonzero entries in each row. In other words, the width of~$\mathbf{A}$ is at most~$L$.

\begin{theorem}
	\label{big equation resolution theorem}
	Let $\mathbf{A} \in \Fp^{m \times n}$ and $\mathbf{b} \in \Fp^{m}$.
	For any $H \subseteq F(\mathbf{\Sigma A}, \mathbf{\Sigma b})$ it holds: $F(\mathbf{A}, \mathbf{b}) \wedge V \lsymd_{\mathcal{O}(m^{\Theta(p)} p^{L+1})+|H|} H$.
\end{theorem}
\begin{proof}
Define $\lambda := \frac{\log(2)}{\log(p/(p-1))}$ and $f(x):= C p^{L+1} x^\lambda$ for some constant~$C$ chosen later. 
Regarding the relationship between~$\lambda$ and~$p$ we have $\lambda \sim \log(2)p$ (i.e.,~$\lim_{n\to \infty} \lambda/\log(2)p=1)$). 

We prove the following by induction over the number of equations $m$: 
	For any $H \subseteq F(\mathbf{\Sigma A}, \mathbf{\Sigma b})$ it holds: $F(\mathbf{A}, \mathbf{b}) \wedge V \lsymd_{f(m)+|H|} H$.

Induction basis: $m=0$. In this case, we have $\mathbf{\Sigma A} = \mathbf{0}$ and $\mathbf{\Sigma b} = 0$. Then $F(\mathbf{\Sigma A}, \mathbf{\Sigma b}) = \emptyset$; hence $H=\emptyset$ and we have nothing to prove.

Induction step: $m-1 \rightarrow m$. Here we have two cases:
\\
Case 1: \emph{Symmetric sum}: 
For this case we assume that for all $\mathbf{d}$ with $\Delta_{\mathbf{d}} \in \mathrm{Sym}(F(\mathbf{\Sigma A}, \mathbf{\Sigma b}))$ we have a $\mathbf{d'}$ with $\mathbf{d}|_{\mathbf{\Sigma A}}=\mathbf{d'}|_{\mathbf{\Sigma A}}$ and $\Delta_{\mathbf{d'}} \in \mathrm{Sym}(F(\mathbf{A}, \mathbf{b}))$.
Thanks to this property, all the symmetries of $F(\mathbf{\Sigma A}, \mathbf{\Sigma b})$ are already present in $F(\mathbf{A}, \mathbf{b})$ and can be used by the symmetry rule. So we only need to derive a few clauses of $F(\mathbf{\Sigma A}, \mathbf{\Sigma b})$ to obtain a set allowing us to generate all clauses via symmetries. This which can be done by inductively applying \autoref{sum resolution} as follows. 

\newcommand{\vhx}[1]{\mathbf{z}^{#1}}
Define $\mathbf{A'}$ and $\mathbf{b'}$ to be the first $m-1$ rows of $\mathbf{A}$ and $\mathbf{b}$ respectively. 
Define $\theta := \lv(\mathbf{\Sigma A'}, \mathbf{A}_{m,\ast})$. We have $|\theta| \leq |\supp(\mathbf{A}_{m,\ast})| \leq L$. If $\mathbf{\Sigma A}\neq\mathbf{0}$, then for each $k\neq \mathbf{\Sigma b}$ there exists a vector $\vhx{k}$ with $\supp(\vhx{k})\subseteq \supp(\mathbf{\Sigma A})$ such that $\mathbf{\Sigma A} \cdot \vhx{k}=k$. Define $G:=\{ C_{\mathbf{\Sigma A}}(\vhx{k}) \mid k\in \Fp\setminus\{\mathbf{\Sigma b} \} \} \subseteq F(\mathbf{\Sigma A},\mathbf{\Sigma b})$. Then $|G|=p-1$.

Using \autoref{sum resolution}, we get $F(\mathbf{\Sigma A'},\mathbf{\Sigma b'})\wedge F(\mathbf{A}_{m,\ast}, \mathbf{b}_{m}) \wedge V \vdash^w_{|G|\cdot\mathcal{O}(p^L)} G$.
This derivation only uses a subset $H' \subseteq F(\mathbf{\Sigma A'},\mathbf{\Sigma b'})$ of at most $|H'| \leq \mathcal{O}(p^{L+1})$ clauses. 
By induction, it holds that $F(\mathbf{A'}, \mathbf{b'}) \wedge V \lsymd_{f(m-1)+|H'|} H'$. We can combine 
these derivations by \autoref{reshelper1} to obtain $F(\mathbf{A}, \mathbf{b}) \wedge V \lsymd_{f(m-1)+\mathcal{O}(p^{L+1})} G$. 
Using $\lambda \geq 1$, we take in total $f(m-1)+C p^{L+1} = C p^{L+1} (m-1)^\lambda + C p^{L+1} \leq C p^{L+1} m^\lambda = f(m)$ steps, for some constant $C$.

Now we show that $G$ is a generator for $H$: Let $c \in F(\mathbf{\Sigma A}, \mathbf{\Sigma b})$. Then $c=C_{\mathbf{\Sigma A}}(\mathbf{x})$ for some $\mathbf{x}\in P(\mathbf{\Sigma A}, \mathbf{\Sigma b})$. Define $\mathbf{d}:= \mathbf{x} - \vhx{\mathbf{\Sigma A \cdot x}}$. Then $\mathbf{\Sigma A \cdot d} = 0$, so $\Delta_{\mathbf{d}}\in \mathrm{Sym}(F(\mathbf{\Sigma A}, \mathbf{\Sigma b}))$. Hence there exists a $\varphi\in \mathrm{Sym}(F(\mathbf{A}, \mathbf{b}))$ such that $\varphi(C_{\mathbf{\Sigma A}}(\vhx{\mathbf{\Sigma A \cdot x}})) = \Delta_{\mathbf{d}}(C_{\mathbf{\Sigma A}}(\vhx{\mathbf{\Sigma A \cdot x}})) = C_{\mathbf{\Sigma A}}(\mathbf{x}) = c$. 
We can apply the local symmetry rule to derive $c$ from $G$ in a single step, using the symmetries of $F(\mathbf{A}, \mathbf{b})$.
Repeating this for every $c\in H$ yields $F(\mathbf{A}, \mathbf{b}) \wedge V \lsymd_{f(m)+|H|} H$.

If $\mathbf{\Sigma A}=\mathbf{0}$ and $\mathbf{\Sigma b}\neq 0$ then $F(\mathbf{\Sigma A}, \mathbf{\Sigma b})=\{ C_{\mathbf{0}}(\mathbf{0}) \} = \{\bot\}=: G$, which can be derived in at most $C p^{L+1}$ steps, again using \autoref{sum resolution}.

If $\mathbf{\Sigma A}=\mathbf{0}$ and $\mathbf{\Sigma b}=0$ then $F(\mathbf{\Sigma A}, \mathbf{\Sigma b})=\emptyset$. We treat this the same way as the case $m=0$.

\medskip
Case 2: \emph{Composite}: 
If Case 1 does not apply, the following must hold by \autoref{symsummatrix}: 
For some $\mathbf{d}$ with $\mathbf{\Sigma A \cdot d}=0$, the equations $\mathbf{A d'} = \mathbf{0}$ and $\mathbf{\diag{\Sigma A} d'} = \mathbf{\diag{\Sigma A} d}$ have no common solution $\mathbf{d'}$.

Applying \autoref{inconsistent lin eq} to the combined inconsistent equations, 
we have $\mathbf{v,w}$ such that $\mathbf{v A}+\mathbf{w \diag{\Sigma A}} = \mathbf{0}$ and $\mathbf{w \diag{\Sigma A} d} \neq 0$.
We will use the vector $\mathbf{v}$ to decompose $\mathbf{A}$ into two smaller matrices, each contributing independently to the derivation of $H$.
First we show that $\mathbf{v}$ has special properties which make this \emph{divide and conquer} approach work. 
Then we need to ensure that the sub-problems are not too large for our proof length bound $f$. 

It holds: $\mathbf{v A} = - \mathbf{w \diag{\Sigma A}}$; thus $\mathbf{v A d} = - \mathbf{w \diag{\Sigma A} d} \neq 0$.
For all $i\in[1,n]$, if $(\mathbf{\Sigma A})_i=0$, then $(\mathbf{v A})_i = (- \mathbf{w \diag{\Sigma A}})_i = - \mathbf{w}_i (\mathbf{\Sigma A})_i  = 0$. Hence $\supp(\mathbf{v A}) \subseteq \supp(\mathbf{\Sigma A})$.
We show that $\mathbf{v A} $ and $\mathbf{\Sigma A}$ are linearly independent:
Let $\alpha_1, \alpha_2 \in \Fp$ such that $\alpha_1\mathbf{v A}+\alpha_2 \mathbf{\Sigma A}=\mathbf{0}$. Then $0 = \alpha_1\mathbf{v A d}+\alpha_2 \mathbf{\Sigma A d} = \alpha_1\mathbf{v A d}$, which implies $\alpha_1=0$. Since $\mathbf{\Sigma A} \neq \mathbf{0}$, we also have $\alpha_2=0$.

Let $k_1 \in \underset{k\in \Fp}{\operatorname{arg\,max}}\,|\{i \mid \mathbf{v}_i=k \}|$ be the most common component of $\mathbf{v}$. 
Let $k_2 \in \underset{k\in \Fp,\, k\neq k_1}{\operatorname{arg\,max}}\,|\{i \mid \mathbf{v}_i=k \}|$ be the second most common component of $\mathbf{v}$.
Since $\mathbf{v A}$ is linearly independent from $\mathbf{\Sigma A}$, we have $\mathbf{v} \neq k \cdot \mathbf{1}$ for all $k\in \Fp$, so there are at least two different components in $\mathbf{v}$. Hence $k_1$ and $k_2$ exist.
Define $m_i$ to be the number of times $k_i$ occurs in~$\mathbf{v}$. We have $m_i \geq 1$ for $i\in\{1,2\}$. Furthermore $m_1 \geq m/p$ and $m_2 \geq (m-m_1)/(p-1)$.

\newcommand{\vi}[1]{\mathbf{v}^{#1}}
Define $\vi{1} := \mathbf{v} - k_1 \mathbf{1}$ and $\vi{2} := k_2 \mathbf{1} - \mathbf{v}$.
It holds: $\vi{1}+\vi{2} = (k_2-k_1)\mathbf{1}$. By subtracting $k_i$ from every component, we get exactly $m_i$ zeros in $\vi{i}$, i.e.\ $|\supp(\vi{i})| = m - m_i$.

Towards a contradiction, assume there is some $j \in \lv(\vi{1} \mathbf{A}, \vi{2} \mathbf{A})$. Then 
$0 = (\vi{1} \mathbf{A}+\vi{2} \mathbf{A})_j = ((k_2-k_1)\mathbf{\Sigma A})_j$, so $0 = (\mathbf{\Sigma A})_j$. Thus $0 = -\mathbf{w}_j (\mathbf{\Sigma A})_j = (\mathbf{v A})_j = (\vi{1} \mathbf{A}+k_1\mathbf{\Sigma A})_j = (\vi{1}\mathbf{ A})_j+k_1(\mathbf{\Sigma A})_j = (\vi{1}\mathbf{A})_j$. This contradicts the assumption. Hence $\lv(\vi{1}\mathbf{A}, \vi{2}\mathbf{A}) = \emptyset$. 
By \autoref{sum resolution} we can derive the sum clauses of $\vi{1}\mathbf{A}+\vi{2}\mathbf{A}$ in 0 steps, so they are already implied by the summand clauses:
$F((\vi{1}+\vi{2})\mathbf{A}, (\vi{1}+\vi{2})\mathbf{b}) \sqsubseteq F(\vi{1}\mathbf{A}, \vi{1}\mathbf{b}) \cup F(\vi{2}\mathbf{A}, \vi{2}\mathbf{b})$.
It follows that
\begin{align*}
F(\mathbf{\Sigma A}, \mathbf{\Sigma b}) &= F((k_2-k_1)\mathbf{\Sigma A}, (k_2-k_1)\mathbf{\Sigma b}) \\
&= F((\vi{1}+\vi{2})\mathbf{A}, (\vi{1}+\vi{2})\mathbf{b}) \\
&\sqsubseteq F(\vi{1}\mathbf{A}, \vi{1}\mathbf{b}) \cup F(\vi{2}\mathbf{A}, \vi{2}\mathbf{b}). 
\end{align*}
Hence we can partition $H \subseteq F(\mathbf{\Sigma A}, \mathbf{\Sigma b})$ into $H_1$ and $H_2$ such that 
$H_i \sqsubseteq F(\vi{i}\mathbf{A}, \vi{i}\mathbf{b})$ for $i\in \{1,2\}$.

Note that $F(\vi{i}\mathbf{A}, \vi{i}\mathbf{b}) = F(\mathbf{\Sigma} \diag{\vi{i}} \mathbf{A}, \mathbf{\Sigma} \diag{\vi{i}} \mathbf{b})$.
Since $|\supp(\vi{i})| \leq m-1$, we have at least one zero row each in $\diag{\vi{1}}\mathbf{A}$ and $\diag{\vi{2}}\mathbf{ A}$. This makes it possible to apply the induction hypothesis, yielding $F(\diag{\vi{i}}\mathbf{A}, \diag{\vi{i}}\mathbf{b}) \wedge V \lsymd_{f(|\supp(\vi{i})|)+|H_i|} H_i$.

Scaling the equations does not produce different clauses, so we have \\ $F(\diag{\vi{1}}\mathbf{A}, \diag{\vi{1}}\mathbf{b}) \cup F(\diag{\vi{2}}\mathbf{A}, \diag{\vi{2}}\mathbf{b}) \subseteq F(\mathbf{A}, \mathbf{b})$. Then we can combine the derivations of $H_1$ and $H_2$ to obtain $F(\mathbf{A}, \mathbf{b}) \wedge V \lsymd_{f(|\supp(\vi{1})|)+f(|\supp(\vi{2})|)+|H|} H$. It holds: 
$f(|\supp(\vi{1})|)+f(|\supp(\vi{2})|) 
= f(m-m_1)+f(m-m_2) 
\leq f(m-m_1)+f(m-(m-m_1)/(p-1)) =: T(m_1)$.
By applying standard calculus techniques to the function $T$, we find that $T(m_1)\leq f(m)$ for all possible values of $m_1$.
\end{proof}

\begin{corollary} \label{lin eq res}
	Let $\mathbf{A} \in \Fp^{m \times n}$ and $\mathbf{b} \in \Fp^{m}$, such that there is no $\mathbf{x} \in \Fp^n$ satisfying $\mathbf{Ax=b}$. Then there exists a resolution refutation of $F(\mathbf{A}, \mathbf{b}) \wedge V$ using the local symmetry rule, with its length bounded by $\mathcal{O}(m^{\Theta(p)} p^{L+1})$.
\end{corollary}

\section{Linear-sized Refutations for Non-Isomorphism of Multipedes}\label{sec:multipedes}
We can use the result on linear equations to show that there are short resolution proofs for the non-isomorphism of Multipede graphs. 

\begin{theorem}\label{thm:multipedes}
	Let $G=(V,W,E)$ be a connected bipartite graph such that $MP(G)$ is asymmetric, and $\omega \in W$. Then $F\left(MP(G)_{a_\omega},MP(G)_{b_\omega}\right)$ has a linear-sized resolution refutation using the local symmetry rule.
\end{theorem}

\label{proof:thm:multipedes}

\begin{proof} 

Let $G=(\{v_1,\dots,v_m\}, \{w_1, \dots, w_n\}, E)$ be a connected bipartite graph 
and $\omega := w_k$ for some $k$. 
Our goal is to apply the techniques of the previous section to the formula $F_0:=F\left(MP(G)_{a_\omega},MP(G)_{b_\omega}\right)$.

We first inspect the simpler formula $F_1:=F(MP(G), MP(G))$. The solutions of this formula correspond to the automorphisms of $MP(G)$. By applying resolution to $F_1$, we can derive the formula $F(\mpmat(G), \mathbf{0})$:
Let $i\in [1,m]$. Then \begin{align*}
F(\mpmat(G)_{i,\ast},0) &= \bigwedge_{\mathbf{x}\in P(\mpmat(G)_{i,\ast}, 0)} \bigvee_{j \in \supp(\mpmat(G)_{i,\ast})} \overline{\xi_{j,\mathbf{x}_j}} \\
&= \bigwedge_{\mathbf{x}\in P(\mpmat(G)_{i,\ast}, 0)} \bigvee_{\{v_i,w_j\}\in E} \overline{\xi_{j,\mathbf{x}_j}} \\
&= \bigwedge_{\substack{B \subseteq \neighborhood{G}(v_i) \\ |B| \text{ odd}}} \left( \bigvee_{w_j \in B} \overline{\xi_{j,1}} \vee \bigvee_{w_j \in \neighborhood{G}(v_i) \setminus B} \overline{\xi_{j,0}} \right)
\end{align*}

Define $N:=\neighborhood{G}(v_i)$. We define $\mathcal{P}_{even}(N)$ to be the subsets of $N$ with even cardinality.
For all $B \subseteq N$ with odd $|B|$ there exists a surjective function $\gamma: \mathcal{P}_{even}(N) \rightarrow N$ such that for all $S\in \mathcal{P}_{even}(N): \gamma(S) \in S\setminus B \cup B \setminus S$.
We can make the following resolution derivation from $F_1$:
\begin{align*}
\text{Type 1: } && \bigvee_{S \in \mathcal{P}_{even}(N)} z^v_{\emptyset, S} \\
\forall S \in \mathcal{P}_{even}(N) \text{ with } w:=\gamma(S)\in B\setminus S: \text{Type 3: } && \overline{z^v_{\emptyset,S}} \vee \overline{y_{a_w,b_w}} \\
\forall S \in \mathcal{P}_{even}(N) \text{ with } w:=\gamma(S)\in S\setminus B: \text{Type 3: } && \overline{z^v_{\emptyset,S}} \vee \overline{y_{a_w,a_w}} \\
\implies && \bigvee_{w \in B} \overline{y_{a_w,b_w}} \vee \bigvee_{w \in N\setminus B} \overline{y_{a_w,a_w}},
\end{align*}
taking $|\mathcal{P}_{even}(N)| \leq 2^{|N|}$ steps. Repeating this process for every $B$ and $i$ takes $\sum_{v\in V} |\mathcal{P}_{odd}(\neighborhood{G}(v))| \cdot 2^{|\neighborhood{G}(v)|} = \mathcal{O}(|F_1|)$ resolution steps. 
Define a variable renaming $r$ on $F(\mpmat(G), \mathbf{0})$ as follows: 
\[r(\xi_{j,\kappa}) := \begin{cases}
y_{a_{w_j},a_{w_j}} &\text{if } \kappa=0\\
y_{a_{w_j},b_{w_j}} &\text{if } \kappa=1
\end{cases}\]
Then we have the derivation $F_1 \vdash_{\mathcal{O}(|F_1|)} r(F(\mpmat(G), \mathbf{0}))$.
As a consequence, $|F(\mpmat(G), \mathbf{0})| = \mathcal{O}(|F_1|)$.
The clauses of $r(V)$
are simply the Type 1 clauses of $F_1$.

To apply \autoref{big equation resolution theorem}, we need to translate the symmetries $\Delta_{\mathbf{d}}\in \mathrm{Sym}(F(\mpmat(G), \mathbf{0}))$ into symmetries of $F_1$. Let $\mathbf{d}\in \mathbb{F}_2^n$ such that $\mpmat(G)\mathbf{d=0}$. Define $D:= \{w_i\in W \mid \mathbf{d}_i=1 \}$. Then the following map $\psi_\mathbf{d}$ is a symmetry of $F_1$: 
	$\text{for }w\in W$ set~$\psi_\mathbf{d}(y_{a_w,a_w})$ to be~$y_{a_w,b_w}$ if~$w\in D$ and~$y_{a_w,a_w}$ otherwise. Similarly~$\psi_\mathbf{d}(y_{a_w,b_w})$ is~$y_{a_w,a_w}$ if~$w\in D$ and~$y_{a_w,b_w}$.
	We also set~$\psi_\mathbf{d}(y_{b_w,a_w})$ to be~$	y_{b_w,b_w}$ if~$w\in D$ and~$y_{b_w,a_w}$ and we set~$\psi_\mathbf{d}(y_{b_w,b_w})$ to be~$	y_{b_w,a_w}$ if~$w\in D$ and~$y_{b_w,b_w}$. Finally for~$v\in V$ we define~$\psi_\mathbf{d}(z^v_{S,T}) := z^v_{S,T \symmdif D}$ 
and have the property $\psi_\mathbf{d}(r(c)) = r(\Delta_{\mathbf{d}}(c))$ for all clauses $c\in F(\mpmat(G), \mathbf{0})$.

Now, if the graph $MP(G)$ is asymmetric, the only solution of $\mpmat(G)\mathbf{y=0}$ is $\mathbf{y=0}$. Then we can deduce $\mathbf{y}_k=0$ from the equation system by combining rows. 
Applying \autoref{big equation resolution theorem}, we get $F(\mpmat(G), \mathbf{0}) \wedge V \lsymd_{\mathcal{O}(m 2^{L+1})} \xi_{k,0}$. 
Renaming variables yields $r(F(\mpmat(G), \mathbf{0})) \wedge r(V) \lsymd_{\mathcal{O}(m 2^{L+1})} y_{a_\omega,a_\omega}$.
As we have seen, $r(F(\mpmat(G), \mathbf{0}))$ and $r(V)$ can be derived from $F_1$ and the symmetries are preserved; hence $F_1 \lsymd_{\mathcal{O}(m 2^{L+1})} y_{a_\omega,a_\omega}$.

Note that $F_0$ is obtained from $F_1$ simply by replacing $y_{a_\omega,a_\omega}$ and $y_{b_\omega,b_\omega}$ with $0$. From this we conclude that, $F_0 \lsymd_{\mathcal{O}(m 2^{L+1})} \bot$. 
\end{proof}

\newpage

\bibliography{library}

\end{document}